\documentstyle[12pt,epsf]{article}

\def\simg{{\ \lower-1.2pt\vbox{\hbox{\rlap{$>$}\lower6pt\vbox{\hbox{$\sim$}}}}\ }}
\def\siml{{\ \lower-1.2pt\vbox{\hbox{\rlap{$<$}\lower6pt\vbox{\hbox{$\sim$}}}}\ }}

\def\als{\alpha_{s}}

\def\lQ{\Lambda_{\rm QCD}}

\newcommand{\MS}{\overline{\rm MS}}
\newcommand{\RS}{\rm RS}
\newcommand{\PS}{\rm PS}
\newcommand{\OS}{\rm OS}

\newcommand{\nn}{\nonumber}
\newcommand{\be}{\begin{equation}}
\newcommand{\ee}{\end{equation}}
\newcommand{\bea}{\begin{eqnarray}}
\newcommand{\eea}{\end{eqnarray}}

\newcommand{\Appendix}[1]%
    {%
     \section{#1}%
      }

\begin{document}\setlength{\unitlength}{1mm}

\begin{titlepage}
\begin{flushright}
\tt{TTP01-12}
\end{flushright}

\vspace{1cm}
\begin{center}
\begin{Large}
{\bf Determination of the bottom quark mass from the
  $\Upsilon(1S)$ system}\\[2cm]
\end{Large} 
{\large Antonio Pineda}\footnote{pineda@particle.uni-karlsruhe.de}\\
{\it Institut f\"ur Theoretische Teilchenphysik\\ 
        Universit\"at Karlsruhe, 
        D-76128 Karlsruhe, Germany }
\end{center}

\vspace{1cm}

\begin{abstract}
  We approximately compute the normalization constant of the first
  infrared renormalon of the pole mass (and the singlet static
  potential).  Estimates of higher order terms in the perturbative
  relation between the pole mass and the $\MS$ mass (and in the
  relation between the singlet static potential and $\als$) are
  given. We define a matching scheme (the renormalon subtracted
  scheme) between QCD and any effective field theory with heavy quarks
  where, besides the usual perturbative matching, the first renormalon
  in the Borel plane of the pole mass is subtracted. A determination
  of the bottom $\MS$ quark mass from the $\Upsilon(1S)$ system is
  performed with this new scheme and the errors studied. Our result
  reads $m_{b,\MS}(m_{b,\MS})=4\,210^{+90}_{-90}({\rm
  theory})^{-25}_{+25}(\als)$ MeV. Using the mass difference between
  the $B$ and $D$ meson, we also obtain a value for the charm quark
  mass: $m_{c,\MS}(m_{c,\MS})=1\,210^{+70}_{-70}({\rm
  theory})^{+65}_{-65}(m_{b,\MS})^{-45}_{+45}(\lambda_1)$ MeV. We
  finally discuss upon eventual improvements of these determinations.
  \vspace{5mm} \\ PACS numbers: 14.65.Fy, 14.65.Dw, 12.38.Cy, 12.39.Hg
\end{abstract}

\end{titlepage}
\vfill
\setcounter{footnote}{0} 
\vspace{1cm}

\section{Introduction}

Systems composed by heavy quarks are very important in the study of
the QCD dynamics. This is due to the fact that they can test QCD in a
kinematical regime otherwise unreachable with only light quarks. They
are characterized by having an scale, the mass of the heavy quark,
much larger than any other dynamical scale in the problem. Therefore,
it seems reasonable to study these systems by using effective field
theories where the mass has been used as an expansion parameter. Some
examples of such are HQET \cite{HQET} for the
the one-heavy quark sector and NRQCD \cite{NRQCD} or pNRQCD
\cite{pNRQCD,long} for the $Q$-$\bar Q$ sector.

\medskip

On general grounds, the matching coefficients of the (QCD) effective
theories suffer from renormalon ambiguities \cite{LMS} (for a review
on renormalons see \cite{renormalons}). This means that, in principle,
one can not compute these matching coefficients with infinity accuracy
in terms of the (short distance physics) parameters of the underlying
theory. From a formal point of view, there is not fundamental problem
related with these ambiguities, since the renormalon ambiguities of
the matching coefficients cancel with the renormalon ambiguities in the
calculation of the matrix elements in the effective theory in such a
way that observables are renormalon free (as they should be) up to the
order in the expansion parameters to which the calculation has been
performed.

Being more specific, a generic matching coefficient $c(\nu/m)$ would have the
following perturbative expansion in $\als$: 
\be
c(\nu/m)=\bar c+\sum_{n=0}^{\infty}c_n\als^{n+1},
\ee
where $\als$ (in the $\MS$ scheme) is normalized at the scale $\nu$. Its Borel transform
would be
\be
B[c](t)\equiv \sum_{n=0}^{\infty}c_n{t^n \over n!},
\ee
and $c$ is written in terms of its Borel transform as 
\be
c=\bar c+\int\limits_0^\infty\mbox{d} t \,e^{-t/\als}
\,B[c](t)
.
\ee
The ambiguity in the matching coefficient reflects in
poles\footnote{In general, this pole becomes a branch point
singularity but this does
not affect the argumentation.} in the
Borel transform. If we take the one closest to the origin, 
\be
\delta B[c](t) \sim {1 \over a-t},
\ee
where $a$ is a positive number, it sets up the maximal accuracy with which one can obtain the matching
coefficients from a perturbative calculation, which is (roughly) of
the order of 
\be
\label{accpert}
\delta c \sim r_{n^*}\als^{n^*},
\ee
where $n^* \sim {a \over\als}$. 
Moreover, the fact that $a$ is
positive means that, even after Borel resummation, $c$ suffers from a
non-perturbative ambiguity of order 
\be
\label{accnp}
\delta c \sim \left(\lQ\right)^{a\beta_0 \over 2\pi}\,. 
\ee

\medskip
\begin{figure}[h]
\hspace{-0.1in}
\epsfxsize=3.8in
\centerline{\epsffile{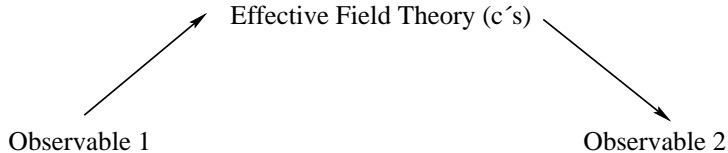}}
\caption {{\it Symbolic relation between observables through the
determination of the matching coefficients of the effective field theory.}}
\label{obs12}
\end{figure}

As we have mentioned, these renormalon ambiguities will cancel anyhow
in the final calculation of the observable. Therefore, we could ask
ourselves why bother about this problem. The answer comes from the
procedure we use to relate (and then predict) observables (see
Fig. \ref{obs12}). Schematically, one takes one observable to fix the
matching coefficient. In order to relate the matching coefficient with
the observable, we use the effective field theory as a tool to provide
the power counting rules in the calculation. Since we are relating one
observable (a renormalon free object) with a matching coefficient
suffering of some renormalon ambiguity, there must be another source
of renormalon ambiguity as to cancel this one. As we have mentioned,
the latter comes from the calculation of the matrix element in the
effective theory. There is a problem here, however. If the matrix
element in the effective theory is a non-perturbative object, the
ambiguity of the matching coefficient is of the same size than this
non-perturbative object that we do not know how to calculate
anyhow. Nevertheless, if the matrix element in the effective theory is
a perturbative object, the only way it has to show the renormalon is
in a bad perturbative behavior in the expansion parameters in the
effective theory (what is happening is that the coefficients that
multiply the expansion parameters are not of $O(1)$ due to the
renormalon, i.e. the renormalon is breaking down the assumption of
naturalness implicit in any effective theory). This may seem
irrelevant since the ambiguity is the same in either case but in the
latter situation it means that the observable is less sensitive to
long distance than the matching coefficient itself. Thus, if, for
instance, we wanted later on to get the short distance parameters from
that (weakly sensitive to long distance physics) observable, we are
not doing an optimal job, since we use an intermediate parameter (the
matching coefficient) that can not be obtained with better accuracy
than the ones displayed in Eqs. (\ref{accpert})-(\ref{accnp}), whereas
the observable (and the short distance parameter) is less sensitive to
long distance physics (the same problem will also appear if we want to
relate two weakly sensitive to long distance physics observables
through a more long distance sensitive matching coefficient). We can
consider three examples (observables) to illustrate this point: the
mass of the meson $B$ ($M_B$), the inclusive semileptonic decay width
of the $B$ ($\Gamma(B \rightarrow X_ul\nu)$) and the mass of
$\Upsilon(1S)$ (assuming in this latter case $m\als^2 \gg \lQ$ for
illustration). For these observables, the first non-perturbative
corrections (leaving aside renormalons) are of the following type: 
\be
\delta_{np} M_B \sim \lQ, \qquad \delta_{np} \Gamma(B \rightarrow
X_ul\nu) \sim G_F^2m_{\OS}^3\lQ^2, \qquad \delta_{np} M_{\Upsilon(1S)}
\sim m_{\OS}{\lQ^4 \over (m_{\OS}\als)^4}\,, 
\ee 
where $m_{\OS}$ is
the pole mass.  The above results only become true if the perturbative
piece can be computed with such precision. Nevertheless, this is
not true in the on-shell (OS) scheme, where one uses the pole mass as
an expansion parameter, since the latter suffers from renormalon
ambiguities \cite{BenekeBraunren,kinetic}. Therefore, effectively, the
above observables can only be computed perturbatively (working with
the pole mass) with the following precision 
\be 
\delta_{np}^{({\rm pert.})}  M_B \sim \lQ, 
\qquad 
\delta_{np}^{({\rm pert.})}  \Gamma (B \rightarrow X_ul\nu)
\sim G_F^2m_{\OS}^4\lQ, 
\qquad 
\delta_{np}^{({\rm pert.})} M_{\Upsilon(1S)} \sim \lQ.  
\ee 
For the first observable this
discussion is irrelevant but not for the other two, which become
rather less well known that they could be.

\medskip

Roughly speaking, the above discussion means that the renormalon of the
matching coefficients can be {\it spurious} (it is not related to a
{\it real} non-perturbative contribution in the observable) or {\it
real} (it is related to a {\it real} non-perturbative contribution
in the observable). In fact, this distinction depends on the
observable we are considering rather than on the renormalon of the
matching coefficient itself\footnote{In the discussion above the
renormalon of the pole mass would be called {\it real} if we think of
the $B$ meson mass and {\it spurious} if we think of the mass of the
$\Upsilon (1S)$.}. The point we want to stress is that at the matching
calculation level it makes no sense this distinction. Therefore, there
is no necessity to keep the renormalon ambiguity (that it only appears
due to the specific factorization prescription we are using) in the
matching coefficients (even more if we take into account that its only
role is to worsen the perturbative expansion in the matching
coefficients). Thus, our proposal is that one should figure out a
matching scheme where the renormalon ambiguity is subtracted from the
matching coefficients. This is the program we will pursue here for the
specific case of effective field theories with heavy quarks. In
principle, the same program should, eventually, be carried out with
other effective theories. Nevertheless, the worsening of the
perturbative expansion was especially evident in effective theories
with heavy quarks. This is due to the fact that the renormalon
singularities lie close together to the origin and that perturbative
calculations have gone very far in this case
\cite{GRA90}-\cite{top}. In other effective field
theories this problem may have not become so acute (yet) but it may
become relevant in the future.

The structure of the paper is as follows. In the next section, we
   compute the normalization constant of the first infrared (IR)
   renormalon of the pole mass. We also give estimates of the higher
   order coefficients of the perturbative series relating the pole
   mass with the $\MS$ mass. In section \ref{secpot}, we compute the
   normalization constant of the first IR renormalon of the singlet static
   potential and also estimates of the higher order perturbative terms
   in the potential are given. In section \ref{secdefRS}, new
   definitions of the pole mass and the singlet static potential are given, within
   an effective field theory perspective, by subtracting their
   closest singularities to the origen in the Borel plane. In section
   \ref{secdetMMS}, we provide a determination of the bottom $\MS$
   mass from the $\Upsilon(1S)$ mass and an estimate of the errors
   within this new approach. In section \ref{secdetmc}, we provide a
   determination of the charm $\MS$ mass by using the mass difference
   between the $B$ and $D$ mesons. In section \ref{conclusions}, we
   give our conclusions and discuss how to improve our
   determinations of the bottom and charm masses.

\section{Mass normalization constant}
\label{secmas}

The pole mass can be related to the $\MS$ renormalized mass -- 
which in principle can be measured to any accuracy at a very high 
energy scale -- by the series
\be
\label{series}
m_{\OS} = m_{\MS} + \sum_{n=0}^\infty r_n \als^{n+1}\,, 
\ee
where the normalization point $\nu=m_{\MS}$ is understood for $m_{\MS}$
(in this way we effectively resum logs that are not associated to the
renormalon since both $m_{\MS}(\nu)$ and $m_{\MS}(m_{\MS})$ do not
suffer from the bad renormalon behavior) and 
the first three coefficients $r_0$, $r_1$ and $r_2$ are known 
\cite{GRA90} ($\als=\als^{(n_l)}$, where $n_l$ is the number of
light fermions; we will assume in this work that any other quark except
the heavy quark is massless). The pole mass is also known to be IR
finite and scheme-independent at any finite order in $\als$ \cite{irfinite}. We then define the Borel transform 
\be\label{borel}
m_{\OS} = m_{\MS} + \int\limits_0^\infty\mbox{d} t \,e^{-t/\als}
\,B[m_{\OS}](t)
\,,
\qquad B[m_{\OS}](t)\equiv \sum_{n=0}^\infty 
r_n \frac{t^n}{n!} . 
\ee
The behavior of the perturbative expansion of
Eq. (\ref{series}) at large
orders is dictated by the closest singularity to the origin of its
Borel transform, which happens to be located at
$t=2\pi/\beta_0$, where we define 
$$
\nu {d \als \over d
\nu}=-2\als\left\{\beta_0{\als \over 4 \pi}+\beta_1\left({\als \over 4
\pi}\right)^2 + \cdots\right\}
.$$  
Being more precise, the behavior of the Borel transform near the
closest singularity at the origin reads (we define $u={\beta_0 t \over 4 \pi}$)
\be
B[m_{\OS}](t(u))=N_m\nu {1 \over
(1-2u)^{1+b}}\left(1+c_1(1-2u)+c_2(1-2u)^2+\cdots \right)+({\rm
analytic\; term}),
\ee
where by {\it analytic term}, we mean a piece that we expect it to be
analytic up to the next renormalon ($u=1$). This dictates the behavior of the perturbative expansion at large orders to be 
\be\label{generalm}
r_n \stackrel{n\rightarrow\infty}{=} N_m\,\nu\,\left({\beta_0 \over 2\pi}\right)^n
\,{\Gamma(n+1+b) \over
\Gamma(1+b)}
\left(
1+\frac{b}{(n+b)}c_1+\frac{b(b-1)}{(n+b)(n+b-1)}c_2+ \cdots
\right).
\ee
The different 
 $b$, $c_1$, $c_2$, etc ... can
 be obtained from the procedure used in \cite{Benb}.
The coefficients $b$ and $c_1$ were computed
(exactly!) in \cite{Benb}, and $c_2$ in \cite{renormalons} (where,
apparently, there are some missprints for its analogous expression $s_2$). They read 
\be
b={\beta_1 \over 2\beta_0^2}\,,
\ee
\be
c_1={1 \over 4\,b\beta_0^3}\left({\beta_1^2 \over \beta_0}-\beta_2\right)
\ee
and 
\be
c_2={1 \over b(b-1)}
{\beta_1^4 + 4 \beta_0^3 \beta_1 \beta_2 - 2 \beta_0 \beta_1^2 \beta_2 + 
   \beta_0^2 (-2 \beta_1^3 + \beta_2^2) - 2 \beta_0^4 \beta_3 
\over 32 \beta_0^8}
\,.
\ee
We then use the idea of \cite{Lee1} (see also \cite{Lee2}) and define the new function
\bea
D_m(u)&=&\sum_{n=0}^{\infty}D_m^{(n)} u^n=(1-2u)^{1+b}B[m_{\OS}](t(u))
\\
\nn
&
=&N_m\nu\left(1+c_1(1-2u)+c_2(1-2u)^2+\cdots
\right)+(1-2u)^{1+b}({\rm analytic\; term})
\,.
\eea
This function is singular but bounded at the first IR renormalon. Therefore, we
can expect to obtain an approximate determination of $N_m$ if we know the first
coefficients of the series in $u$ and by using 
\be
N_m\nu=D_m(u=1/2)
.
\ee
The first three coefficients: $D_m^{(0)}$, $D_m^{(1)}$ and $D_m^{(2)}$
are known in our case. In order the calculation to make sense, we
choose $\nu \sim m$. In other words, we avoid to have another large (small) parameter, otherwise we should dealt with the necessity
of resummation. If, for illustration, we restrict ourselves to the
large $\beta_0$ approximation, the dimensionful parameters rearrange
in the quantity 
$$
m\left({\nu \over m}\right)^{2u} \simeq \nu\{1+(2u-1)\ln{\nu \over m}+\cdots\}
.
$$
Therefore, the underlying assumption is that we are in a regime where
(besides $2u-1 \ll 1$)
$$
(2u-1)\ln{\nu \over m} \ll 1\,.
$$
For the specific choice $\nu=m$, we obtain (up to $O(u^3)|_{u=1/2}$)
\bea
\label{nm}
N_m&=&0.424413+0.137858+0.0127029= 0.574974 \quad (n_f=3)
\\
\nn
&=&0.424413+0.127505+0.000360952= 0.552279 \quad (n_f=4)
\\
\nn
&=&0.424413+0.119930-0.0207998= 0.523543 \quad (n_f=5)
\eea
\medskip
\begin{figure}[h]
\hspace{-0.1in}
\epsfxsize=4.8in
\centerline{\epsffile{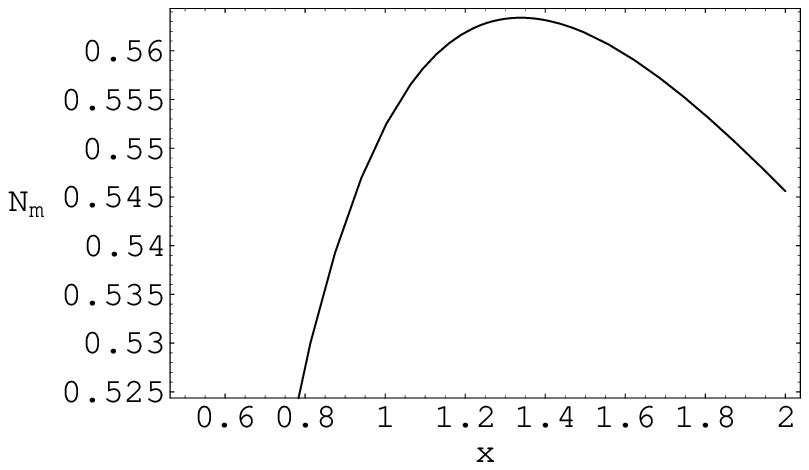}}
\caption {{\it $x \equiv {\nu \over m_{\MS}}$ dependence of
  $N_m$ for $n_f=4$.}} 
\label{figNm}
\end{figure}
The convergence is surprisingly good. One can also see that the scale
dependence is quite mild (there even appears a place of minimal
sensitive to the scale dependence, see Fig. \ref{figNm}). If, for
illustration, we take ${\nu \over m_{\MS}}=2$ one obtains, for $n_f=4$,
$N_m=0.545580$. On the other hand, for small ${\nu \over m_{\MS}}$, the scale dependence
starts to become important (if, for illustration, we take ${\nu \over
  m_{\MS}}=1/2$, one obtains, for $n_f=4$, $N_m=0.434619$).

\medskip

By using Eq. (\ref{generalm}), we can now go backwards and give some estimates for the $r_n$. They are displayed in Table \ref{tabm}. We can
see that they go closer to the exact values of $r_n$ when increasing
$n$. This makes us feel confident that we are near the asymptotic
regime dominated by the first IR renormalon and that for higher $n$ our
predictions will become an accurate estimate of the exact values. In
fact, they are quite compatible with the results obtained by other
methods like the large $\beta_0$ approximation (see Table \ref{tabm}).
If we also compare with the estimates for $r_3$ given in Ref. \cite{rn},
our results are also roughly compatible with those up to differences of
the order of 
10 \%, 20 \% and 30 \% for $n_f=3$, $n_f=4$ and $n_f=5$, respectively.

\begin{table}[h]
\addtolength{\arraycolsep}{0.2cm}
$$
\begin{array}{|l||c|c|c|c|c|}
\hline
 {\tilde r}_n=r_n/m_{\MS}  & {\tilde r}_0 & {\tilde r}_1 & {\tilde r}_2 & {\tilde r}_3 & {\tilde r}_4  
\\ \hline\hline
{\rm exact}\; (n_f=3) & 0.424413 & 1.04556 & 3.75086  & --- &
 ---  \\
{\rm Eq.}\; (\ref{generalm})\; (n_f=3) & 0.617148  & 0.977493 &  3.76832
  & 18.6697  &  118.441  \\
{\rm large}\; \beta_0\; (n_f=3) & 0.424413  & 1.42442 &  3.83641
  & 17.1286   &  97.5872   \\ \hline\hline
{\rm exact}\; (n_f=4) & 0.424413 & 0.940051 & 3.03854 & --- 
 & ---   \\
{\rm Eq.}\; (\ref{generalm})\; (n_f=4) & 0.645181 & 0.848362 & 3.03913
 & 13.8151 & 80.5776   \\ 
{\rm large}\; \beta_0\; (n_f=4) & 0.424413 & 1.31891 &  3.28911
  & 13.5972  & 71.7295    \\ \hline\hline
{\rm exact}\; (n_f=5) &0.424413 & 0.834538  & 2.36832 & --- & ---  \\
{\rm Eq.}\; (\ref{generalm})\; (n_f=5) & 0.706913 & 0.713994 & 2.36440 &
 9.73117  & 51.5952 \\
{\rm large}\; \beta_0\; (n_f=5) & 0.424413  & 1.21339 & 2.78390 
  & 10.5880  & 51.3865   \\  \hline
\end{array}
$$
\caption{{\it Values of $r_n$ for $\nu=m_{\MS}$. Either the exact result
(when available), the estimate using Eq. (\ref{generalm}), or the estimate using the large
  $\beta_0$ approximation \cite{BenekeBraun}.}}
\label{tabm}
\end{table}

In our estimates in Table \ref{tabm}, we have included
(formally) subleading terms in the $1/n$ expansion up to $O(1/n^2)$. We show
their impact in Table \ref{tab1n}. They happen to be corrections with
respect the leading order result in all cases and of the order of 4 \%
or smaller for $n \simg 2$. The smallness of the $O(1/n)$ corrections
with respect the leading order result can be traced back to the
approximate (numerical?) pattern $\beta_n \sim \beta_0^n$ (for $\beta_1$
and $\beta_2$) for $n_f=0$ and that $n_f$ terms do not jeopardize this
rule (this explains the strong dependence of the results with $n_f$). In
fact, if this rule were exact, one can easily see that $c_n=0$ for all
$n$. The breaking of this rule explains the non-zero values for $c_n$.
The $O(1/n^2)$ terms depend on $\beta_3$, which has been recently
evaluated in \cite{runningbeta}, through $c_2$.  In this case, however,
we rather have $\beta_3 \sim 2 \beta_0^3$ and the $n_f$ terms are quite
large breaking the pattern above. In fact, $c_1$ (the upper entry for
$r_0$ in Table \ref{tab1n}) strongly depends on $n_f$ so that if for
$n_f=0$, $c_1$ and $c_2$ (the lower entry for
$r_0$ in Table \ref{tab1n}) are of the same size ($c_1 \simeq -0.215$, $c_2 \simeq 0.185$), for $n_f=5$, $c_1$
becomes an order of magnitude smaller than $c_2$. Therefore, one may
think of the smallness of $c_1$ as a numerical accident (for some
specific values of $n_f$) not reflecting the natural size of the
$O(1/n)$ terms and therefore explaining the apparent breakdown (or slow
convergence) of the $1/n$ expansion. This would be elucidated if
$\beta_4$ were known. In the mean time, we will stick to this belief.

\begin{table}[h]
\addtolength{\arraycolsep}{0.2cm}
$$
\begin{array}{|l||c|c|c|c|c|}
\hline
 {\tilde r}_n=r_n/m_{\MS}  & {\tilde r}_0 & {\tilde r}_1 & {\tilde r}_2 & {\tilde r}_3 & {\tilde r}_4  
\\ \hline\hline
O(1/n)\; (n_f=3) & -0.164 & -0.046 & -0.027  &-0.019 & -0.015
\\
O(1/n^2)\; (n_f=3) & 0.237  & -0.103 & -0.017 &-0.007 & -0.004  
\\ \hline\hline
O(1/n)\; (n_f=4) & -0.105 & -0.028 & -0.016 &-0.012  & -0.009   
\\
O(1/n^2)\; (n_f=4) &0.274 & -0.126 &-0.020 &-0.008 &  -0.004
\\ \hline\hline
O(1/n)\;(n_f=5) &0.024 & 0.006 & 0.003 & 0.002 & 0.002  
\\
O(1/n^2)\; (n_f=5) & 0.326 & -0.165 &-0.023 & -0.009 &  -0.005  
\\  \hline
\end{array}
$$
\caption{{\it $O(1/n)$ corrections (normalized with respect the leading solution)
    of our $r_n$ estimates for different number of light fermions.}}
\label{tab1n}
\end{table}

We can now try to see how the large $\beta_0$ approximation works in
the determination of $N_m$. In order to do so, we study the one chain
approximation from which we obtain the value \cite{BenekeBraunren} 
\be
N_m^{({\rm large}\; \beta_0)}={C_f \over \pi}e^{5 \over 6}=0.976564.
\ee 
By comparing with Eq. (\ref{nm}), we can see that it does not
provide an accurate determination of $N_m$. This may seem to be in
contradiction with the accurate values that the large $\beta_0$
approximation provides for the $r_n$ (starting at $n=2$) in Table
\ref{tabm}.  Lacking of any physical explanation for this fact, it
may just be considered to be a numerical accident. In fact, the agreement
between our determination and the large $\beta_0$ results does not
hold at very high orders in the perturbative expansion, whereas we
believe, on physical grounds, since our approach incorporates the
exact nature of the renormalon, that our determination should go
closer to the exact result at high orders in perturbation
theory. Nevertheless, the large $\beta_0$ approximation remains
accurate up to relative high orders.

\section{Static singlet potential normalization constant}
\label{secpot}

One can think of playing the same game with the singlet 
static potential in the situation where $\lQ \ll 1/r$. The potential, however, is not an IR
safe object at any order in the perturbative expansion
\cite{Appelquist,short}. Its perturbative expansion reads 
\be
V_s^{(0)}(r;\nu_{us})=\sum_{n=0}^\infty V_{s,n}^{(0)} \als^{n+1},
\ee
where we have made explicit its dependence in the IR cutoff $\nu_{us}$. The first
three coefficients $V_{s,0}^{(0)}$, $V_{s,1}^{(0)}$ and $V_{s,2}^{(0)}$ are known
\cite{FSP} as well as the leading-log terms of
$V_{s,3}^{(0)}$ \cite{short} (for the renormalization-group improved expression see
\cite{RGPS}). Nevertheless, these leading logs are not
associated to the first IR renormalon, since they also appear in momentum
space (see also the discussion below), so
we will not consider them further in this section (or if one prefers, we
will only consider coefficients that we fully know). 

We now use the observation that the first IR renormalon of the singlet
 static potential cancels with the renormalon of (twice) the pole
 mass. This has been proven in the (one-chain) large $\beta_0$
 approximation in \cite{thesis,long} and at any loop (disregarding
 eventual effects due to $\nu_{us}$) in \cite{BenekePS}. We would like
 to argue that this cancellation holds without resort to any
 diagrammatic analysis as far as factorization between the different
 scales in the physical system is achieved. This can be done within an
 effective field theory framework where any renormalon ambiguity
 should cancel between operators and matching coefficients. Therefore,
 in the situation where $1/r \gg \lQ$, one can do the matching between
 NRQCD and pNRQCD and $2m_{\OS}+V_s^{(0)}$ can be understood as an
 observable up to $O(r^2\lQ^3,\lQ^2/m)$ renormalon (and/or
 non-perturbative) contributions (see
 Eq. (\ref{pnrqcdph}))\footnote{Note that the same argumentation does
 not apply for the octet static potential $V_o^{(0)}$. The reason is
 that, even at leading order in $1/m$, $2m_{\OS}+V_o^{(0)}$ is not an
 observable. This is due to the fact that there is still interaction
 with low energy gluons, as one can see from
 Eq. (\ref{pnrqcdph}). Therefore, one expects $2m_{\OS}+V_o^{(0)}$ to
 be ambiguous by an amount of $O(\lQ)$.}. This would prove the (first
 IR) renormalon cancellation at any loop (as well as proving the
 independence of this IR renormalon of $\nu_{us}$). In a way, the
 argumentation would be similar to the one made for the renormalon
 cancellation in the electromagnetic correlator using the operator
 product expansion \cite{BenekeBraunren,kinetic}, where the renormalons
 are absorbed in local condensates. In our case, however, we are
 talking of non-local objects, but fortunately, effective field
 theories provide themselves as useful in these situations.

We can now read the asymptotic behavior of the static potential from the one
 of the pole mass and work analogously to the
 previous section. We define the Borel transform 
\be\label{borelb}
V_s^{(0)} = \int\limits_0^\infty\mbox{d} t \,e^{-t/\als}\,B[V_s^{(0)}](t)
\,,
\qquad 
B[V_s^{(0)}](t)\equiv \sum_{n=0}^\infty 
V_{s,n}^{(0)} \frac{t^n}{n!} . 
\ee
The closest singularity to the origen is located at 
$t=2\pi/\beta_0$. This dictates the behavior of the perturbative expansion at large orders to be 
\be\label{generalV}
V_{s,n}^{(0)} \stackrel{n\rightarrow\infty}{=} N_V\,\nu\,\left({\beta_0 \over 2\pi}\right)^n
 \,{\Gamma(n+1+b) \over
 \Gamma(1+b)}
\left(
1+\frac{b}{(n+b)}c_1+\frac{b(b-1)}{(n+b)(n+b-1)}c_2+ \cdots
\right)
,
\ee
and the Borel transform near the singularity reads
\be
B[V_{s}^{(0)}](t(u))=N_V\nu {1 \over
(1-2u)^{1+b}}\left(1+c_1(1-2u)+c_2(1-2u)^2+\cdots \right)+({\rm
analytic\; term}).
\ee
In this case, by {\it analytic term}, we mean an analytic function up to
 the next IR renormalon at $u=3/2$ \cite{Aglietti}. 

\begin{figure}[h]
\hspace{-0.1in}
\epsfxsize=4.8in
\centerline{\epsffile{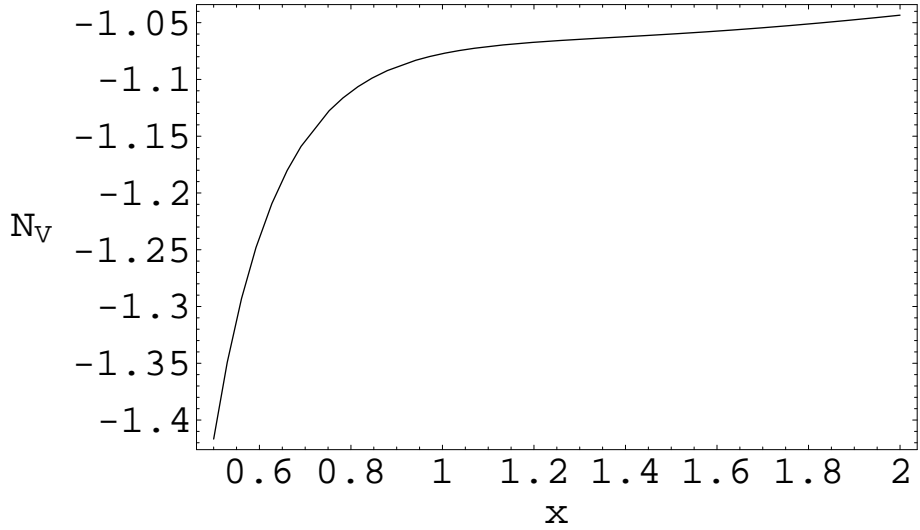}}
\caption {{\it $x \equiv {\nu r}$ dependence of
  $N_V$ for $n_f=4$.}} 
\label{figNV}
\end{figure}

As in the previous section, we define the new function
\bea
D_V(u)&=&\sum_{n=0}^{\infty}D_V^{(n)} u^n = (1-2u)^{1+b}B[V_{s}^{(0)}](t(u))
\\
\nn
&
=&N_V\nu\left(1+c_1(1-2u)+c_2(1-2u)^2+\cdots
\right)+(1-2u)^{1+b}({\rm analytic\; term})
\eea
and try to obtain an approximate determination of $N_V$ by using the
 first three (known) coefficients of this series. By a discussion
 analogous to the one in the previous section, we fix $\nu=1/r$. We
 obtain (up to $O(u^3)|_{u=1/2}$)
\bea
\label{nv}
N_V&=&-1.33333+0.571943-0.345222 = -1.10661 \quad (n_f=3) \\ \nn
&=&-1.33333+0.585401-0.329356 = -1.07729 \quad (n_f=4) \\ \nn
&=&-1.33333+0.586817-0.295238 = -1.04175 \quad (n_f=5) \,.  
\eea 
The
convergence is not as good as in the previous section. Nevertheless,
it is quite acceptable and, in this case, apparently, we have a sign
alternating series. In fact, the scale dependence is quite mild (see
Fig. \ref{figNV}) except (again) for small values of $\nu r$. Overall,
up to small differences, the same picture than for $N_m$ applies.

\medskip

So far we have not made use of the fact that $2N_m+N_V = 0$. We use
this equality as a check of the reliability of our calculation. We can see
that the cancellation is quite dramatic. We obtain
\begin{eqnarray*} 
2{2N_m+N_V \over 2N_m-N_V} = \,\left\{
\begin{array}{ll}
\displaystyle{ 0.038 }&\ \ , \, n_f=3 \\
\displaystyle{0.025}&\ \ ,\, n_f=4 \\
\displaystyle{0.005}&\ \ ,\, n_f=5.
\end{array} \right.
\end{eqnarray*}
We should stress that the evaluation of $N_m$ and $N_V$ uses
independent inputs. Therefore, this cancellation seems to
be nontrivial. 
This makes us confident that the number we have obtained for $N_m$ (or
$N_V$) is quite accurate. The difference is even better than what one would
have expected from the last terms in the series of $N_V$, but in this case,
this may due to the fact that the series is sign-alternating. 

In the following we will use the determination of $N_m$ to fix $N_V$. 
Any difference should be included in the errors.

\medskip

We can now obtain estimates for $V_{s,n}^{(0)}$ by using Eq.
(\ref{generalV}).  They are displayed in Table \ref{tabv}. Note that in
Table \ref{tabv} no input from the static potential has been used since
even $N_V$ have been fixed by using the equality $2N_m=-N_V$. We can see
that the exact results are reproduced fairly well (the same discussion
than for the $r_n$ determination applies). This makes us feel confident
that we are near the asymptotic regime dominated by the first IR
renormalon and that for higher $n$ our predictions will become an
accurate prediction of the exact results. The
comparison with the values obtained with the large $\beta_0$
approximation would go (roughly) along the same lines than for the mass
case, although the large $\beta_0$ results seem to be less accurate in
this case (see Table \ref{tabv}).

\begin{table}[h]
\addtolength{\arraycolsep}{0.2cm}
$$
\begin{array}{|l||c|c|c|c|c|}
\hline
{\tilde V}_{s,n}^{(0)}= r V_{s,n}^{(0)}  & {\tilde V}_{s,0}^{(0)} & {\tilde V}_{s,1}^{(0)} & {\tilde V}_{s,2}^{(0)} & {\tilde V}_{s,3}^{(0)} & {\tilde V}_{s,4}^{(0)} 
\\ \hline\hline
{\rm exact}\; (n_f=3) & -1.33333 & -1.84512 & -7.28304  & --- &
 ---  \\
{\rm Eq.}\; (\ref{generalV})\; (n_f=3) & -1.23430 & -1.95499 & -7.53665
  & -37.3395  & -236.882  \\
{\rm large}\; \beta_0\; (n_f=3) & -1.33333  & -2.69395 & -7.69303
  & -34.0562   &  ---  \\
 \hline\hline
{\rm exact}\; (n_f=4) & -1.33333 & -1.64557 & -5.94978 & --- 
 & ---   \\
{\rm Eq.}\; (\ref{generalV})\; (n_f=4) & -1.29036 & -1.69672 & -6.07826
 & -27.6301  & -161.155   \\ 
{\rm large}\; \beta_0\; (n_f=4) & -1.33333 & -2.49440  &  -6.59553
  & -27.0349  & ---  \\ 
 \hline\hline
{\rm exact}\; (n_f=5) & -1.33333 & -1.44602 & -4.70095 & --- & ---  \\
{\rm Eq.}\; (\ref{generalV})\; (n_f=5) &-1.41383 & -1.42799 & -4.72881 &
  -19.4623 & -103.190  \\
{\rm large}\; \beta_0\; (n_f=5) & -1.33333  &-2.29485  &  -5.58246
  & -21.0518  & --- \\
 \hline
\end{array}
$$
\caption{{\it Values of $V_{s,n}^{(0)}$ with $\nu=1/r$. Either the exact result (when available), the
  estimate using Eq. (\ref{generalV}), or the estimate using the large
  $\beta_0$ approximation \cite{Sumino1,Hoangcharm1}.}}
\label{tabv}
\end{table}

In order to avoid large corrections from terms depending on $\nu_{us}$,
the predictions should be understood with $\nu_{us}=1/r$ and later on
one can use the renormalization group equations for the static potential
\cite{RGPS} to keep track of the $\nu_{us}$ dependence.

Finally, we would also like to mention that the previous discussion
about the determination of $N_m$ in the large $\beta_0$ approximation
also applies here for the determination of $N_V$.

\section{Renormalon subtracted matching and power counting}
\label{secdefRS}
 
In effective theories with heavy quarks, the inverse of the heavy
quark mass becomes one of the expansion parameters (and matching
coefficients). A natural choice in the past (within the infinitely
many possible definitions of the mass) has been the pole mass because
it is the natural definition in OS processes where the particles
finally measured in the detectors correspond to the fields in
the Lagrangian (as in QED). Unfortunately, this is not the case in QCD
and one reflection of this fact is that the pole mass suffers from renormalon singularities. Moreover, these renormalon singularities lie
close together to the origin and perturbative calculations have gone
very far for systems with heavy quarks. At the practical level, this
has reflected in the worsening of the perturbative expansion in
processes where the pole mass was used as an expansion parameter
\cite{PY,sumrules1}. It
is then natural to try to define a new expansion parameter replacing
the pole mass but still being an adequate definition for threshold
problems. This idea is not new and has already been pursued in the
literature, where several definitions have arisen. For instance, the
kinetic mass \cite{kinetic}, the PS mass \cite{BenekePS}, the 1S mass
\cite{Hoang1S} and the ${\overline {\rm PS}}$ mass
\cite{overlinePS}. We can not resist the tentation of trying our own
definition. We believe that, having a different systematics than the
other definitions, it could further help to estimate the errors in the
more recent determinations of the $\MS$ quark mass. Our definition, as the
definitions above, try to cancel the bad perturbative behavior
associated to the renormalon. On the other hand, we would like to
understand this problem within an effective field theory
perspective. From this point of view what one is seeing is that the
coefficients multiplying the (small) expansion parameters in the
effective theory calculation are not of
natural size (of $O(1)$). The natural answer to this problem is that
we are not properly separating scales in our effective theory and some
effects from small scales are incorporated in the matching
coefficients. These small scales are dynamically generated in $n$-loop
calculations ($n$ being large) and are of $O(m\,e^{-n})$ (we are
having in mind a large $\beta_0$ evaluation) producing the bad
(renormalon associated) perturbative behavior. The natural way to
deal with this problem would be to perform an expansion of the (small)
scale $m\,e^{-n}$ over the (large) scale $m$, or alike, for a
$n$-loop calculation. Unfortunately, this is something that standard
dimensional regularization does not know how to achieve since it does
not know how to separate the scale $m$ from the scale $m\,e^{-n}$
treating them on the same footing\footnote{This problem does not
appear in hard cut-off renormalization schemes or alike (to which,
after all, will belong our scheme), which explicitely cut-off these
scales. At this respect, we can not avoid thinking that we are in a
similar situation to the beginnings of NRQCD (NRQED), where it was
better known how to achieve the separation of scales with hard cut-off
than with dimensional regularization \cite{NRQCD,Labelle}. This
problem was first solved in \cite{Manohar,pNRQCD} within an effective
field theory framework (see also \cite{BenekeSmirnov} where a solution
within a diagrammatic approach was provided). A key point in the
solution came by realizing that the way to implement the separation of
scales (matching) in dimensional regularization was by first expanding
the Feynman integrals with respect the small scales (the ones that
would be kept in the effective theory) prior to
integration. Obviously, a similar solution here (with no necessity of
hard cutoff), where one should expand with respect the small scale
$me^{-n}$ prior to integration, would be most welcome. This
would change the perturbative expansion and the (small) scale
$me^{-n}$ should be kept explicit in the effective theory. To date, we
are not able to further substantiate this discussion but we expect to
come back to this issue in the future.}. In order to overcome this
problem, we may think of doing the Borel transform. In that case, the
renormalon singularities correspond to the non-analytic terms in
$1-2u$. These terms also exist in the effective theory.  Therefore,
our procedure will be to subtract the pure renormalon contribution in
the new mass definition, which we will call renormalon subtracted
(RS) mass, $m_{\RS}$ (with no pretentious aims, all the other mass
definitions do cancel the renormalon as well, but rather for
notational purposes).  We define the Borel transform of $m_{\RS}$ as
follows 
\be 
B[m_{\RS}]\equiv B[m_{\OS}] -N_m\nu_f {1 \over
(1-2u)^{1+b}}\left(1+c_1(1-2u)+c_2(1-2u)^2+\cdots \right), 
\ee 
where
$\nu_f$ could be understood as a factorization scale between QCD and
NRQCD and, at this stage, should be smaller than $m$. The
expression for $m_{\RS}$ reads 
\be
\label{mrsvsmpole}
m_{\RS}(\nu_f)=m_{\OS}-\sum_{n=0}^\infty  N_m\,\nu_f\,\left({\beta_0 \over
2\pi}\right
)^n \als^{n+1}(\nu_f)\,\sum_{k=0}^\infty c_k{\Gamma(n+1+b-k) \over
\Gamma(1+b-k)}
\,,
\ee
where $c_0=1$. 
We expect that with this renormalon free definition the 
coefficients multiplying the expansion parameters in the effective
theory calculation will have a natural size and also the coefficients multiplying
the powers of $\als$ in the perturbative expansion relating $m_{\RS}$ with $m_{\MS}$. Therefore,
we do not loose accuracy if we first obtain $m_{\RS}$ and later on we
use the perturbative relation between $m_{\RS}$ and
$m_{\MS}$ in order to obtain the latter. Nevertheless, since we will work order by order in
$\als$ in the relation between $m_{\RS}$ and
$m_{\MS}$, it is important to expand everything in terms of
$\als$, in particular $\als(\nu_f)$, 
in order to achieve the renormalon cancellation order by order in
$\als$. Then, the
perturbative expansion in terms of the $\MS$ mass reads 
\be
m_{\RS}(\nu_f)=m_{\MS} + \sum_{n=0}^\infty r^{\RS}_n\als^{n+1}\,,
\ee
where $r^{\RS}_n=r^{\RS}_n(m_{\MS},\nu,\nu_f)$. These $r^{\RS}_n$ are
the ones expected to be of natural
size (or at least not to be artificially enlarged by the first IR renormalon).
 
From Eq. (\ref{mrsvsmpole}), we can see that we are subtracting a
finite piece (that should correspond to the renormalon for large $n$)
for every $r_n$. It is more than debatable whether we should do any subtraction
for $r_0$. Therefore, to test the scheme dependence of our results, we also define a modified RS scheme, RS', as follows
\be 
m_{RS'}(\nu_f)=m_{\OS}-\sum_{n=1}^\infty  N_m\,\nu_f\,\left({\beta_0 \over
2\pi}\right
)^n \als^{n+1}(\nu_f)\,\sum_{k=0}^\infty c_k{\Gamma(n+1+b-k) \over
\Gamma(1+b-k)}
\,,
\ee
and the Borel transform corresponds to
\be
B[m_{RS'}]\equiv B[m_{RS}] +N_m\nu_f \left(1+c_1+c_2+\cdots \right)
\,. 
\ee 

At this stage, we would like to make some preliminary numbers to estimate the effect
of our definition as well as to compare with other threshold masses. In order to
simplify the discussion we define an static version of the 1S mass:
\be
m_{1S}^{({\rm static})}\equiv m_{\OS}+{V(r) \over
2}=m_{\MS}+\left(r_0-{C_f \over 2r}\right)\als+\cdots\,,
\ee
and compare the expansions for the pole
mass, the PS mass, the RS mass and the (static) 1S mass. We display
the results in Table \ref{tabmassesnuf2} for the bottom quark and in
Table \ref{tabmassestop} for
the top quark. By default the value $\als(M_z)=0.118$ is
understood. The four loop evolution equation \cite{runningbeta} has been used for the
running of $\als$ as 
provided by the program RunDec.m \cite{RunDec}. Overall, all the threshold mass
definitions work well in the cancellation of the renormalon, as we can
see by comparing with the pole mass result at higher orders. For the
bottom quark case, we have displayed results for $\nu_f=2$ and
$\nu_f=1$ since on the one hand, on conceptual grounds, we would like
to keep $\nu_f$ below (or of the order of) what it will (roughly) be the typical scales of 
the inverse Bohr radius in the $\Upsilon(1S)$ system but, on the other
hand, we would like to keep it significantly larger than $\lQ$. This
gives us little room to play. For the 1S (static mass), we have chosen,
for a closer comparison with the other definitions, $1/r=2,1$ GeV,
respectively. For the top quark case, we have chosen $\nu_f=1/r=3$ GeV. 
\begin{table}[h]
\addtolength{\arraycolsep}{0.2cm}
$$
\begin{array}{|l||c|c|c|c|c|}
\hline
 {\rm Masses}  & O(\als) & O(\als^2) & O(\als^3) & O(\als^4)
 & {\rm total}  
\\ \hline\hline
m_{\OS} &401  & 199 & 144 & 147 & 5\,102  \\ 
\hline \hline
m_{\RS} & 111 & 50  & 17 & 7  & 4\,395   \\
m_{\RS'}& 401 & 114 & 38 & 15 & 4.778   \\ 
m_{\PS}& 210  & 80  & 42  & ---   & 4\,542    \\ 
m_{1S}^{({\rm static})}& 102  & 50  & 19 & 8   & 4\,389    \\ 
\hline \hline
m_{\RS} & 256 & 95 & 40 & 21  & 4\,622  \\
m_{\RS'}& 401 & 157 & 74 & 41  & 4.882   \\ 
m_{\PS}& 306 & 120 & 67  & ---   & 4.703   \\ 
m_{1S}^{({\rm static})}& 251  & 94 & 41 & 22   & 4\,619    \\ 
\hline
\end{array}
$$
\caption{{\it Contributions at various orders in $\als$ for different mass
definitions for the bottom quark case, either with $\nu_f=1/r=2$ GeV
(middle panel) or with $\nu_f=1/r=1$ GeV (lower panel). The results
are displayed in MeV. For the
$O(\als^4)$ results, the estimate from Table
\ref{tabm} has been used. The other parameters have been fixed to the values
$m_{\MS}(m_{\MS})=4.21$ GeV, $\nu=m_{\MS}(m_{\MS})$ and $n_f=4$.}}
\label{tabmassesnuf2}
\end{table}
\begin{table}[h]
\addtolength{\arraycolsep}{0.2cm}
$$
\begin{array}{|l||c|c|c|c|c|}
\hline
 {\rm Masses}  & O(\als) & O(\als^2) & O(\als^3) & O(\als^4)
 & {\rm total}  
\\ \hline\hline
m_{\OS} & 7.585  & 1.615  & 0.497 & 0.221 & 174.917  \\ 
\hline \hline
m_{\RS} & 7.355  &  1.469  & 0.391 & 0.139  & 174.354   \\
m_{\RS'}& 7.585 & 1.590 & 0.461 & 0.180 & 174.816   \\ 
m_{\PS} & 7.447 & 1.518  & 0.460  & ---   & 174.425    \\ 
m_{1S}^{({\rm static})}& 7.368  & 1.475 & 0.395 & 0.141   & 174.379    \\ 
\hline
\end{array}
$$
\caption{{\it Contributions at various orders in $\als$ for different mass
  definitions for the top quark case. The results
are displayed in GeV. For the $O(\als^4)$ results, the estimate from
Table \ref{tabm} has been used. We have used the following set of
parameters: $\nu_f=1/r=3$ GeV,
$m_{\MS}(m_{\MS})=165$ GeV, $\nu=m_{\MS}(m_{\MS})$ and $n_f=5$.}}
\label{tabmassestop}
\end{table}

\medskip

The shift from the pole mass to the RS mass  affects the explicit 
expression of the effective Lagrangians. In particular, in HQET, at
leading order, a residual mass term appears in the Lagrangian
\be
{\cal L}=\bar h \left(iD_0-\delta m_{\RS}\right)h+O\left({1 \over
m_{RS}}\right) \,,
\ee
where $\delta m_{\RS}=m_{\OS}-m_{\RS}$ and similarly for the NRQCD
Lagrangian.  

\medskip

For heavy quark--antiquark systems in the situation where $\lQ \ll
m\als$, it is convenient to integrate out the soft scale ($\sim m\als$) in
NRQCD ending up in pNRQCD. If we consider the leading order in $1/m$, the residual mass term is absorbed in the
static potential (in going from NRQCD to pNRQCD, one runs down the scale
$\nu_f$ up to $\nu_f \siml m\als$). We can then,
analogously to the RS mass, define an singlet static RS (RS')
potential 
\be
V^{(0)}_{s,\RS(\RS')}(\nu_f)=V^{(0)}_s+2\delta m_{\RS(\RS')}
\,,
\ee
where the
coefficients multiplying the perturbative series should be of $O(1)$
(provided that we expand $V^{(0)}_s$ and $\delta m_{RS}$ in the same parameter,
namely $\als$). Notice also the trivial fact that the scheme dependence
of $m_{\RS}$ cancels with the scheme dependence of $V_{RS}$. It is
interesting to see the impact of this definition in the improvement of
the perturbative expansion in the potential. Then, following analogously the discussion
for the masses, we have compared our definition (the
RS potential) with the PS potential and the singlet static potential for some
typical values appearing in the
bottom and top quark case. We have displayed the results in Tables
\ref{tabpotentialbottom} and \ref{tabpotentialtop}. We see that the
improvement is quite dramatic with respect the expansion in the
singlet static
potential, yet, again, for the bottom case, we have little room for
changing $\nu_f$.
\begin{table}[h]
\addtolength{\arraycolsep}{0.2cm}
$$
\begin{array}{|l||c|c|c|c|c|}
\hline
 {\rm Potentials}  & O(\als) & O(\als^2) & O(\als^3) & O(\als^4)
 & {\rm total}  
\\ \hline\hline
V^{(0)}_s & -910  & -306  & -302 & -383 & -1\,902 \\ 
\hline \hline
V^{(0)}_{s,\RS} & -205 & 3  & -2 & -3 & -208  \\
V^{(0)}_{s,\RS'}& -910 &-54  & -14 & -6 & -984 \\ 
V^{(0)}_{s,\PS} & -446  & -42  & -25  & ---   & -513    \\  
\hline \hline
V^{(0)}_{s,\RS} & -558  & -63  &-41  & -26  & -687  \\
V^{(0)}_{s,\RS'}& -910 & -180 & -95 & -54 & -1\,239 \\ 
V^{(0)}_{s,\PS} & -678  & -116  & -75  & ---   & -869 \\  
\hline
\end{array}
$$
\caption{{\it Contributions at various orders in $\als$ for different
singlet static potential 
definitions for some typical scales in the $\Upsilon$ system, either with $\nu_f=2$ GeV
(middle panel) or with $\nu_f=1$ GeV (lower panel). The results are
displayed in MeV. For the
$O(\als^4)$ results, the estimate from Table
\ref{tabv} has been used. The other parameters have been fixed to the values
$\nu=1/r=2.5$ GeV and $n_f=4$.}}
\label{tabpotentialbottom}
\end{table}
\begin{table}[h]
\addtolength{\arraycolsep}{0.2cm}
$$
\begin{array}{|l||c|c|c|c|c|}
\hline
 {\rm Potentials}  & O(\als) & O(\als^2) & O(\als^3) & O(\als^4)
 & {\rm total}  
\\ \hline\hline
V^{(0)}_s & -5\,679  & -874  &-404 &-237 &-7\,194  \\ 
\hline \hline
V^{(0)}_{s,\RS} & -5\,077 & -548  & -185  & -67 & -5\,876  \\
V^{(0)}_{s,\RS'}& -5\,679 & -788  & -294  & -120 & -6\,881  \\ 
V^{(0)}_{s,\PS} & -5\,317 & -648  & -233  & ---  & -6\,199  \\  
\hline
\end{array}
$$
\caption{{\it Contributions at various orders in $\als$ for different
singlet static potential 
definitions for some typical scales in $t$-$\bar t$ systems near threshold. The results are
displayed in MeV. For the
$O(\als^4)$ results, the estimate from Table
\ref{tabv} has been used. The parameters have been fixed to the values
$\nu=1/r=30$ GeV, $\nu_f=3$ GeV and $n_f=5$.}}
\label{tabpotentialtop}
\end{table}

The pNRQCD Lagrangian in the RS scheme formally reads equal than in
the OS scheme:
\begin{eqnarray}                         
& & {\rm L}_{\rm pNRQCD} = \int d^3{\bf R}d^3{\bf r} \Biggl[
{\rm Tr} \,\Biggl\{ {\rm S}^\dagger \left( i\partial_0 
- {{\bf p}^2\over m_{\RS}} +{{\bf p}^4\over 4m_{\RS}^3}
- V^{(0)}_{s,{\RS}}(r) - {V_{s,{\RS}}^{(1)} \over m_{\RS}}- {V_{s,{\RS}}^{(2)} \over m_{\RS}^2}+ \dots  \right) {\rm S}
\nonumber \\
&& \nonumber 
\qquad \qquad + {\rm O}^\dagger \left( iD_0 - {{\bf p}^2\over m_{\RS}}
- V^{(0)}_{o,{\RS}}(r) 
+\dots  \right) {\rm O} \Biggr\}
\nonumber\\
& &\qquad + g V_A ( r) {\rm Tr} \left\{  {\rm O}^\dagger {\bf r} \cdot {\bf E} \,{\rm S}
+ {\rm S}^\dagger {\bf r} \cdot {\bf E} \,{\rm O} \right\} 
+ g {V_B (r) \over 2} {\rm Tr} \left\{  {\rm O}^\dagger {\bf r} \cdot {\bf E} \, {\rm O} 
+ {\rm O}^\dagger {\rm O} {\bf r} \cdot {\bf E}  \right\}
\Biggr]  
\nonumber\\
& &\qquad-\int d^3{\bf R} {1\over 4} G_{\mu \nu}^{a} G^{\mu \nu \, a},
\label{pnrqcdph}
\end{eqnarray}
where $V^{(0)}_{o,{\RS}}=V^{(0)}_o+2\delta m_{\RS(\RS')}$ and the
$1/m$ potentials also get (straightforwardly) affected by rewriting
the expansion in $1/m_{OS}$ in terms of $1/m_{RS}$ (see
\cite{pNRQCD,long,LL} for the definitions in the OS scheme and details). 
The above Lagrangian provides the appropriate description of systems
for which, for their typical $r$, one has the inequality $1/r \gg \lQ$. If
one further assumes that $r^2\lQ^3 \ll m\als^2$, the power counting
rules tell us that the leading solution corresponds to a Coulomb-type
bound state being the non-perturbative effects corrections with
respect to this leading solution. We will assume to be in this situation
in the following. 

One of the ultraviolet cutoffs in pNRQCD is $\nu_{us}$, which cutoffs
the three-momentum of the gluons in pNRQCD and fulfills the relation
$m\als^2 \ll \nu_{us} \ll m\als$. Without any further assumption about the
(perturbative or non-perturbative) behavior at scales below
$\nu_{us}$, one can compute the heavy quarkonium
spectrum. Formally, we have the following expression (at the
practical level, one could work in the OS scheme and
do the replacement to the RS scheme, with the proper power counting,
at the end)  
\be
\label{Mnlj}
M_{nlj}=2m_{\RS}+\sum_{m=2}^{\infty}A_{nlj}^{m,\RS}(\nu_{us})\als^m+\delta M_{nlj}^{\rm US}(\nu_{us})
\,,
\ee
where the $\nu_{us}$ scale dependence of the different pieces
cancels in the overall sum (for the perturbative sum this dependence first appears
in  $A_{nlj}^{5,\RS}$). 

We expect that by working with the RS scheme (or with any other
achieving the renormalon cancellation) the coefficients multiplying
the powers of $\als$ will now be of natural size and therefore the
convergence improved compared with the OS scheme. So far, the
$A_{nlj}^{n,\RS}$ coefficients are exactly known for $n=2,3,4$ whereas
some partial information is also known for  $A_{nlj}^{5,\RS}$ and
$\delta M_{nlj}^{(US)}$. We will discuss them further in the next
section. 

\medskip

At this stage, we would like to discuss some few theoretical issues
that may appear in the readers mind with respect these investigations
(see also the discussion in Ref. \cite{Benekereport}). First, once one
agrees to give up using the pole mass as an expansion parameter, one
may still wonder why not to use the $\MS$ mass instead. There are
several answers to this question. If we consider pNRQCD, working with
$m_{\MS}$ would mean introducing a large shift\footnote{This is
certainly so for $t$-$\bar t$ physics. Nevertheless, for the bottom,
the $O(m\als)$ term does not seem to be that large numerically (see
Table \ref{tabmassesnuf2}), being much smaller than the typical values
of the soft scale in the $\Upsilon(1S)$. Therefore, it may happen that
working with the $\MS$ mass does not destroy the power counting rules
of pNRQCD (or HQET) at the practical level. However, it does from a
conceptual point of view, in particular, one would have problems to
assign power counting rules. Moreover, it would not incorporate the
expected physical fact that for scales between $m$ and $\nu_f \gg \lQ$
the threshold masses are the proper expansion parameters for
processes near on-shell.}, of $O(m\als)$, in the pNRQCD Lagrangian,
therefore jeopardizing the power counting rules. In fact, the same
happens in HQET. Thus, it seems necessary to work first in an scheme
where the power counting rules are preserved, this invalidates the
$m_{\MS}$ mass (as far as we are talking about processes where the
heavy quarks are near OS). On the other hand, we also require not to
have large (not natural) coefficients multiplying the perturbative
expansion. This invalidates the pole mass, remaining only the
threshold schemes. Nevertheless, once one observable is obtained in a
(safe) threshold scheme, one could consider to rewrite it in terms of
the $\MS$ mass. Even that could be eventually dangerous. If, for
instance, we consider the quantity $M_{nlj}$ and expand it in terms of
$m_{\MS}$, there appear, in principle, two problems.  One has to do
with the resummation of logs, which, after all, is one of the
motivations of the whole factorization program between
different scales that effective field theories are (another is to
provide, in an easy way, the power counting rules of the dynamics). By
expanding everything in terms of $\als$, we reintroduce a potentially
large log, $\ln{m \over \nu}$, in the coefficients multiplying the
powers of $\als$ (note that we can not minimize this log if at the
price of introducing another large log, $\ln{m\als \over
\nu}$). Another problem is that, due to the fact that there is another
scale, $m\als$, besides $m$, at least on conceptual grounds, we would
not achieve the renormalon cancellation order by order in $\als$ but
it would occur between different orders in $\als$ jeopardizing, in
principle, the convergence of the perturbative expansion.

\medskip

From the above discussion, we have seen that it is crucial to have a
renormalon free expansion parameter {\it and} an scheme that preserves
the power counting rules or, equivalently, to be able to write an
effective Lagrangian (with the dynamics, i.e.  the power counting
rules) in terms of the new threshold masses. This is something that it
is achieved by the threshold schemes, i.e. the {\it kinetic}, the
PS-like, the 1S and the RS schemes. However, the 1S scheme seems to
rely on assuming that the $Q$-$\bar Q$ system is a mainly perturbative
system. Note also that the 1S and PS schemes depend on $\nu_{us}$. We
see that the power counting rules in the RS scheme are the same than
the ones in the OS scheme, the difference being that it is now
expected that the terms in the perturbative expansion will be of
natural size. We have then been able to solve the renormalon problem
without giving up the factorization between different scales that is
provided by effective field theories (in the OS scheme). Note also
that the RS mass only knows of scales above the cut-off of the
effective theory, a desirable feature in any factorization program.

\medskip

Throughout this work much emphasis has been put on working with
effective field theories. Therefore, one may honestly ask whether the
problems with renormalons here exposed would disappear if one gives up
using effective field theories. Nevertheless, a closer inspection
seems to show that the renormalon problem always appears in physical problems
where one has different scales and wants to achieve factorization
between them. Thus, any procedure (name it
effective field theory or not) that provides factorization and power
counting rules will have the same kind of problems.

\section{Bottom $\MS$ quark mass determination}
\label{secdetMMS}
\begin{figure}[h]
\hspace{-0.1in}
\epsfxsize=4.8in
\centerline{\epsffile{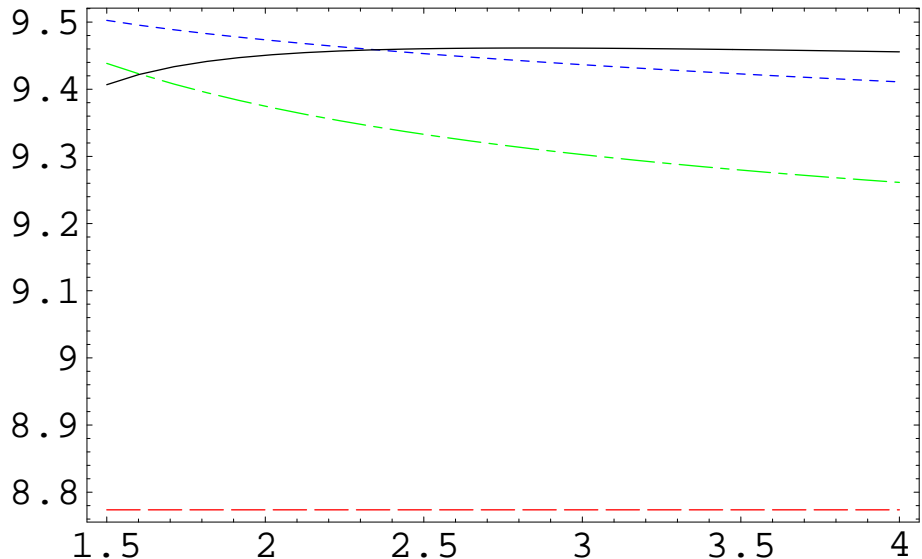}}
\caption{{\it We plot $2m_{b,\RS}$ (dashed line), and the LO (dot-dashed line), NLO (dotted line) and NNLO
  (solid line) predictions for the
  $\Upsilon(1S)$ mass in terms of $\nu$ in the RS scheme. The value of
  $m_{b,\RS}$ is taken from Eq. (\ref{MRSdet}).} }
\label{figM1SRS}
\end{figure}

\begin{figure}[h]
\hspace{-0.1in}
\epsfxsize=4.8in
\centerline{\epsffile{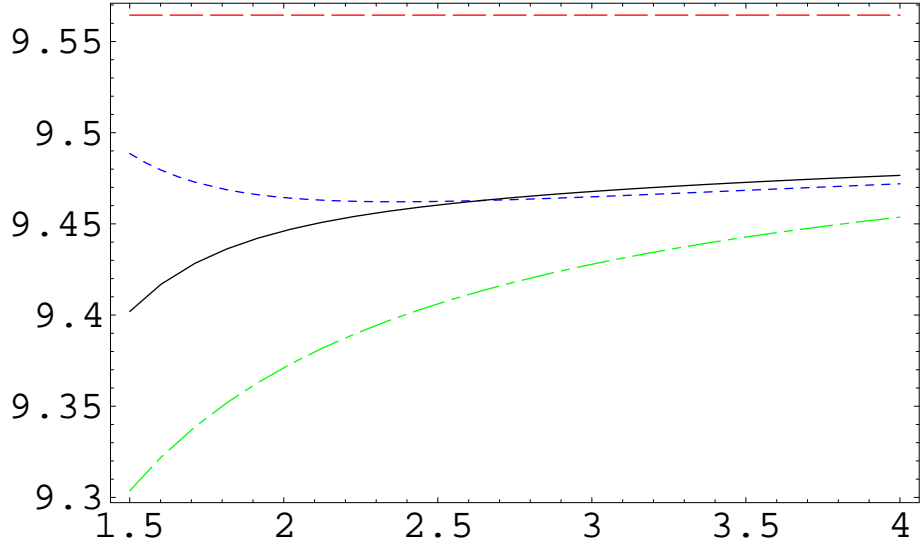}}
\caption{{\it We plot $2m_{b,\RS'}$ (dashed line), and the LO (dot-dashed
line), NLO (dotted line) and NNLO (solid line) predictions for the
  $\Upsilon(1S)$ mass in terms of $\nu$ in the RS' scheme. The value of
  $m_{b,\RS'}$ is taken from Eq. (\ref{MRSprimedet}).} }
\label{figM1SRSprime}
\end{figure}

In this section, we will determine the $\MS$ (and RS) bottom mass from
the $\Upsilon(1S)$ mass. We will use the known results at
next-to-next-to-leading order (NNLO) in the OS scheme \cite{PY} and
rewrite them in the RS scheme where the coefficients multiplying the
perturbative expansion are expected to be of a natural size (for the
moment we neglect ultrasoft contributions). In Figures \ref{figM1SRS}
and \ref{figM1SRSprime}, we plot the scale dependence of the LO, NLO
and NNLO predictions for the $\Upsilon(1S)$ mass in the RS and RS' scheme.


We can also check the dependence of our results with respect to the
theoretical and experimental parameters. As a first estimation of the
errors, we allow for a variation of $\nu$, $\nu_f$, $\als$ and
$N_m$ as follows: $\nu=2.5^{+1.5}_{-1}$ GeV, $\nu_f=2\pm 1$ GeV,
  $\als(M_z)=0.118\pm 0.003$ and $N_m=0.552\pm 0.0552$. For the RS
scheme, we obtain the following result\footnote{Here and in the following,
in the determination of $m_{\MS}$, we have used our estimate of the four-loop relation.}
\be
\label{MRSdet}
m_{b,\RS}(2\;{\rm GeV})=4\,387^{+2}_{+28}(\nu)^{-5}_{+7}(\nu_f)^{-16}_{+16}
(\als)^{-68}_{+68}(N_m)\;{\rm MeV};
\ee
\be
\label{MMSRSdet}
m_{b,\MS}(m_{b,\MS})=4\,203^{+2}_{+25}(\nu)^{-5}_{+6}
(\nu_f)^{-28}_{+27}(\als)^{-10}_{+10}(N_m)\;{\rm MeV}.
\ee
For the RS' scheme, we obtain the result (with the same variation of the
parameters)
 \be
\label{MRSprimedet}
m_{b,\RS'}(2\;{\rm GeV})=4\,782^{-08}_{+31}(\nu)^{-7}_{+3}(\nu_f)
^{+15}_{-12}(\als)^{-28}_{+28}(N_m)\;{\rm MeV};
\ee
\be
\label{MMSRSprimedet}
m_{b,\MS}(m_{b,\MS})=4\,214^{-08}_{+28}(\nu)^{-6}_{+3}
(\nu_f)^{-25}_{+25}(\als)^{-9}_{+9}(N_m)\;{\rm MeV}.
\ee
The $\MS$ mass result depends weakly on the variation of $\nu_f$ and
$N_m$. For the latter it may seem surprising in view of the strong
dependence of the RS mass but, to some extent, the variation of $N_m$
can be understood as a change of scheme.

With the results obtained in Eqs. (\ref{MRSdet})
and (\ref{MRSprimedet}), the expansion for the $\Upsilon(1S)$ mass
would read in the RS scheme (see Eq. (\ref{Mnlj})): 
\be
\label{mupsilonRS}
M_{\Upsilon(1S)}=8\,774+559+120+7\; {\rm MeV},
\ee
and in the RS' scheme:
\be
\label{mupsilonRSprime}
M_{\Upsilon(1S)}=9\,564-158+56-2\; {\rm MeV}\,.
\ee
Both series seem to show convergence. Nevertheless, only in the RS'
scheme a more physical interpretation can be given where the leading
order solution corresponds to the {\it negative} Coulomb binding
energy (but we can see that this affects very little the determination
of the $m_{\MS}$ mass). 

\begin{figure}[h]
\hspace{-0.1in}
\epsfxsize=4.8in
\centerline{\epsffile{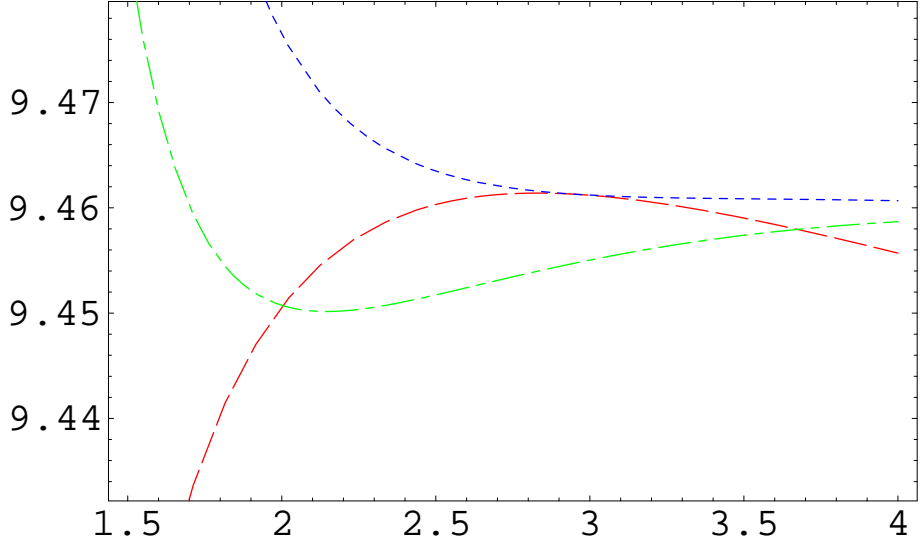}}
\caption{{\it We plot the NNLO (dashed line) versus the NNNLO
estimates in the RS scheme. a) NNNLO estimate without hard/ultrasoft
logs (dot-dashed line) and b) NNNLO estimate including hard/ultrasoft
logs with $\nu_{us}=1$ GeV (dotted line). The value of $m_{b,\RS}$ is
taken from Eq. (\ref{MRSdet}).} }
\label{figRSLL}
\end{figure}

\begin{figure}[h]
\hspace{-0.1in}
\epsfxsize=4.8in
\centerline{\epsffile{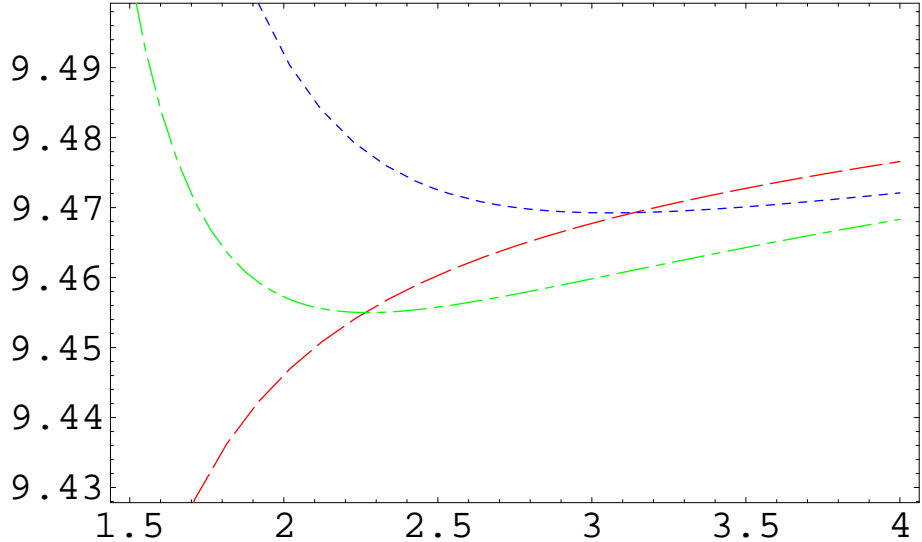}}
\caption{{\it We plot the NNLO (dashed line) versus the NNNLO
estimates in the RS' scheme. a) NNNLO estimate without hard/ultrasoft
logs (dot-dashed line) and b) NNNLO estimate including hard/ultrasoft
logs with $\nu_{us}=1$ GeV (dotted line). The value of $m_{b,\RS'}$ is
taken from Eq. (\ref{MRSprimedet}).} }
\label{figRSprimeLL}
\end{figure}

Naively, from Eqs. (\ref{mupsilonRS})-(\ref{mupsilonRSprime}), one
would be very optimistic about the errors. Nevertheless, let us try to
go deeper into the error analysis.  First, it may be not realistic to
conclude from these results that the magnitude of the NNLO terms is of
order of few MeV. The relative size between different orders will
depend on the scale (as one can see in Figs. \ref{figM1SRS} and
\ref{figM1SRSprime}). The major problems come at small $\nu$ and we
will discuss them later. Leaving them aside, for the RS' scheme, the
NNLO contributions remain quite small, whereas for RS scheme they can
be of the order of $-50$ MeV for large values of $\nu$. Being even more
conservative, one can split the NNLO contributions into the ones due
to the singlet static potential ($+62$ MeV and $45$ MeV, with $\nu=2.5$ GeV,
for the RS and RS' scheme, respectively) and the relativistic ($-55$
MeV and $-47$ MeV, with $\nu=2.5$ GeV, for the RS and RS' scheme,
respectively) ones. Each of them happen to be of the order of 50 MeV
with opposite signs. As far as the relativistic corrections is concerned, we can see that
they are comfortably smaller than the $O(m\als^2)$ leading term. For
the correction due to the singlet static potential, convergence is also found in
the RS scheme, whereas in the RS' scheme the convergence is quite
slow. In fact, this is due to the fact that the magnitude of the LO
and NLO terms in the RS scheme is larger than in the RS' 
scheme. Nevertheless, the discussion on the magnitude of the
corrections due to the static potential is quite delicate since it 
strongly depends on the value of $\nu_f$ used.  A closer look also
shows that the main part of the scale dependence comes from the
relativistic corrections, which become very important at small scales.
Nevertheless, one can think of this scale dependence arising because
of neglecting some higher order logs that, on theoretical grounds, we
know anyway. The, $O(m_{\OS}\als^5)$,
NNNLO logs are known in the OS scheme \cite{LL} (see also
\cite{KP,HMS}). They may be 
considered to be of two different origen. Either those which can be
minimized by choosing $\nu$ of the order of the inverse Bohr radius (for
these an explicit expression can be found in Ref. \cite{Sumino1}), or
those related either to the hard $(\sim
\ln{m}$), or to the ultrasoft scale $(\sim \ln{\nu_{us}}$). Therefore, the only piece
of $A_{101}^{5,\OS}$ left unknown (although by far the most difficult one)
is the log-independent one. For this, there exists, at least, a large
$\beta_0$ determination \cite{Sumino1,Hoangcharm1}. In order, this
evaluation in the OS scheme to be useful, one has to change to the RS
scheme, where the renormalon cancellation is explicit and the NNNLO
correction is expected to be of natural size (note that one has to
take into account the change from the pole mass to the RS mass in the
expressions). Once a value for the log-independent term of
$A_{101}^{5,\OS}$ is assumed (we will use the large $\beta_0$ result), one
can compute $A_{101}^{5,\RS}$ without any further ambiguity and, thus,
the complete NNNLO correction. We will use this result to give an
estimate of the NNNLO corrections (and, therefore, further
substantiate our previous discussion on the convergence of the series), and in order to study the scale
dependence. Note that at this stage a dependence on $\nu_{us}$
appears. If we believe that the large $\beta_0$ approximation provides
(numerically) a good estimate of, at least, the size of the renormalon
contribution to the binding energy in the OS scheme, we could expect
our determination to provide a rough estimate of higher order
effects. We show our results in Figs. \ref{figRSLL} and
\ref{figRSprimeLL}.

Let us first concentrate on the scale dependence. We first consider
  the result without the inclusion of the logs related to the
  hard/ultrasoft scale (being them of different physical origen). The
  results are the dot-dashed line in Figs. \ref{figRSLL} and
  \ref{figRSprimeLL}. At small scales, we see the typical oscillatory
  behavior between different orders in the perturbative expansion
  when the series ends to converge. For scales where the calculation
  is reliable, we see that the correction goes in the direction one
  would expect by choosing a somewhat smaller value for $\nu$ (and
  therefore closer to the scale $m_{\RS}C_f\als \sim 2$ GeV) in the
  NNLO evaluation. On the other hand, if we are concerned about the
  absolute size of the NNNLO corrections, what we can see is that,
  according to our estimate, they are small. We can now add the large
  logs associated to the hard and ultrasoft scale (we use Eq. (19)
  from Ref. \cite{LL}). They depend on $\nu_{us}$. We have chosen
  $\nu_{us}=1$ GeV. Our results correspond to the dot lines in
  Figs. \ref{figRSLL} and \ref{figRSprimeLL}. We see that the
  correction is relatively small (and it goes closer to the dot-dashed
  result when we increase $\nu_{us}$). It also blows up for small
  values of $\nu$ (this makes us believe that in order to properly
  deal with the scale dependence at small $\nu$ a renormalization-group
  improved evaluation could be useful). Therefore, we feel relatively
  confident about these sources of NNNLO corrections. However, this 
  evaluation can not give an estimate of {\it genuine}
  $O(m_{\RS}\als^5)$ corrections coming from other sources. We will discuss
  some of them right now as well as non-perturbative effects.

\medskip

So far, we have just considered the expansion in
$A_{nlj}^{n,\RS}\als^n$ and neglected the ultrasoft effects. Let us
consider them now. An explicit expression for the leading correction
due to ultrasoft effects can be obtained using the multipole expansion
without any assumption about the perturbative or non-perturbative
behavior at ultrasoft scales. It reads \cite{Vol,LL} (we actually
write the Euclidean expression for easier comparison with other
results in the literature)
\begin{equation}
\delta M_{nlj}^{\rm US}(\nu_{us}) \simeq \delta M_{nl}^{\rm US} (\nu_{us}) = {T_F \over 3 N_c}  \int_0^\infty \!\! dt 
\langle n,l |{\bf r} e^{-t(H_o^{\RS}-E_n^{\RS})} {\bf r}| n,l \rangle \langle g{\bf E}^a(t) 
\phi(t,0)^{\rm adj}_{ab} g{\bf E}^b(0) \rangle(\nu_{us}), 
\label{energyUS}
\end{equation}
where $H_o^{\RS} \equiv \displaystyle{{\bf p}^2\over m_{\RS}} +{1
\over 2N_c}{\als \over r}$
and $E_n^{\RS} \equiv - m_{\RS}C_f^2\als^2/(4n^2)$. Different possibilities appear depending on the relative size of $\lQ$ with
respect to the ultrasoft scale $m\als^2$. Let us first study whether
the $\Upsilon(1S)$ lives in the situation where $\lQ \ll m\als^2$. In this case,
the scale $m\als^2$ can be treated perturbatively and one can lowers
further the cutoff up to the situation $\lQ \ll \nu'_{us} \ll
m\als^2$. Scales below $\nu'_{us}$ can be parameterized in terms of
local condensates within an expansion in $O({\lQ^2 \over m^2\als^4})$. Eq. (\ref{energyUS}) would then read (with leading-log
accuracy)
\be
\label{energyUSsmall}
\delta M_{nl}^{\rm US} (\nu_{us}) = 
\delta M_{nl}^{\rm US,\,pert.}+
\delta M_{nl}^{\rm US,\,no-pert.}
\,,
\ee
\bea
\label{energyUSsmallpert}
\delta M_{nl}^{\rm US,\,pert.} &=&
E_n^{\RS} {\als^3 \over \pi} \ln{\nu_{us} \over m_{\RS}\als}
\left\{{C_A \over 3} \left[{C_A^2 \over 2} 
+4C_AC_f {1\over n(2l+1)}+2C_f^2\left({8 \over n(2l+1)} - {1 \over n^2}\right) \right]
 \right.
\nn \\ 
\nn && 
\left.
\qquad +{ C_f^2\delta_{l0} \over 3 n}8\left[C_f-{C_A \over 2}\right]
\right\}
\,,
\eea
and
\be
\label{energyUSsmallnopert}
\delta M_{nl}^{\rm US,\,no-pert.}=\sum_{n=0}^{\infty}C_nO_n
\,,
\ee
where $C_n \sim 1/(m_{\RS}^{3+2n}\als^{4+4n})$ and $O_n \sim \lQ^{4+2n}$ (see \cite{nlonp} for
details). $C_0O_0$ was first computed in
Refs. \cite{Leut} and $C_1O_1$ in Ref. \cite{nlonp}. 
The perturbative leading-log $\nu_{us}$ scale dependence would cancel against the
scale dependence of $A_{10j}^{5,\RS}$ producing $\log(\als)$-type terms.
The validity of these results  for the physical system under 
study rely on two assumptions that one should check. One the one hand,
the final result will depend on $\als(m\als^2)$ (this is easily seen
within an effective field theory framework) but $m\als^2$ is an
small scale for the $\Upsilon(1S)$. If we take as an estimate $\langle H_o-E_1 \rangle_{10} \sim 360$ MeV, we see that
perturbation theory is not truthsworthy for this scale (although,
obviously, numerical factors can play a role). If one, anyway, wants
to study this situation, one can include the log-dependent terms from
Eq. (\ref{energyUSsmall}) in the NNNLO perturbative estimate (the
dependence on $\nu_{us}$ effectively disappears). This would produce
a $\sim +50$ MeV shift in the RS scheme ($\sim +59$ MeV in the RS'
scheme) with respect to the NNLO results (for $\nu=2.5$ GeV), 
becoming even larger for small scales but decreasing for larger values
of $\nu$ (in any case the shift depends on finite piece prescriptions) with,
somewhat, the same shape than the dotted lines in Figs. \ref{figRSLL}
and \ref{figRSprimeLL}. On the other hand, as
far as the non-perturbative corrections is concerned, one should check
that the operator product expansion in terms of local condensates 
converges. If we use our central values (in the RS' scheme) plus the
central values for the condensates used in Refs. \cite{PY,nlonp}, we
obtain
\be
C_0O_0+C_1O_1=144-143\;{\rm MeV}
\,.
\ee
We see that we do not find convergence. The situation improves by
lowering the scale $\nu$ in $\als$ ($97-66$ MeV for $\nu=2$ GeV and
$53-21$ MeV for $\nu=1.5$ GeV) and it also depends on the values
of the condensates, which are poorly known\footnote{The difference with the conclusions in Ref. \cite{nlonp,PY} follows
from the fact that in these works ${\tilde \als}$, as defined in
\cite{PY}, was used as the expansion parameter instead of $\als$ used here. Unfortunately, ${\tilde \als}$ suffers from the renormalon
ambiguity making it potentially large and bringing convergence to the
operator product expansion.}.  
We see that this working hypothesis is not favored by the central
set of parameters of the $\Upsilon(1S)$ system, although it can not be
ruled out if one scan over the possible values of the 
parameters (in fact, if one considers $\nu=2$ GeV a more natural scale
for the soft scale, the operator product expansion would be on the verge of convergence). Nevertheless, for $t$-$\bar t$ production near threshold, the 
situation $\lQ \ll m\als^2$ may be applicable and the results
explained above useful.

\medskip

It may then seem that the $\Upsilon(1S)$ system lives
in the situation where $\lQ \sim m\als^2$. In that case, we can not
lower further $\nu_{us}$ and the chromoelectric correlator in 
Eq. (\ref{energyUS}) cannot be computed using perturbation theory. We
are then faced with the necessity of computing a non-perturbative
non-local condensate. If in the previous case the knowledge of the local
condensates was poor the situation is even worse now. This non-local
condensate is related to the gluonic
correlator for which the most general parameterization reads \cite{Dosch1}
\bea
\langle gF_{\mu\nu}^a(x) 
\phi(x,0)^{\rm adj}_{ab} gF_{\rho\sigma}^b(0) \rangle
&=&
(\delta_{\mu\rho}\delta_{\nu\sigma} - 
\delta_{\mu\sigma}\delta_{\nu\rho})
\left[ {\cal D}(x^2) + {\cal D}_1(x^2) \right]  \nonumber \\
& +& (x_\nu x_\sigma \delta_{\mu\rho} - x_\nu x_\rho \delta_{\mu\sigma} 
- x_\mu x_\sigma \delta_{\nu\rho} + x_\mu x_\rho \delta_{\nu\sigma})
{\partial{\cal D}_1(x^2) \over \partial x^2} ,
\eea
where for the gauge string a straight line is understood and ${\cal D}$ and ${\cal D}_1$ are invariant functions of $x^2$. 
In our case the following combination appears\footnote{At this stage, we
  would like to report some discrepancies with the  results of \cite{Simonov1}. If
  we consider the situation $mv \gg \lQ \gg mv^2$, the leading non-perturbative
  effects can be parameterized in terms of a potential term as follows
  \cite{Balitsky,long}
\begin{equation}
\delta V = r^ 2{T_F \over 3 N_c}  \int_0^\infty \!\! dt  \langle g{\bf E}^a(t) 
\phi(t,0)^{\rm adj}_{ab} g{\bf E}^b(0) \rangle(\nu_{us}). 
\label{potentialNP}
\end{equation}
In Ref. \cite{Simonov1}, the following result was reported: 
\begin{equation}
\delta V = r^ 2{T_F \over N_c} \int_0^\infty \!\! dt  
\left({\cal D}(t^2) +{1 \over 2} {\cal D}_1(t^2)\right).
\label{potentialNPsimonov}
\end{equation}
We have difficulties to
accommodate this result with Eq. (\ref{potentialNP}) after using
Eq. (\ref{EEcorr}). In particular, Eq. (\ref{potentialNPsimonov}) does
not appear to be able to reproduce the leading logs predicted by
perturbation theory.} 
\be
\label{EEcorr}
\langle g{\bf E}^a(t) 
\phi(t,0)^{\rm adj}_{ab} g{\bf E}^b(0) \rangle
=
3\left({\cal D}(t^2) + {\cal D}_1(t^2)+t^2{\partial{\cal D}_1(t^2)
\over \partial t^2} \right).
\ee
In Ref. \cite{DiGiacomo}, a lattice evaluation of the gluonic correlator
was performed. The following parameterization was used to describe
the lattice data (with $t > 0$)
\be
\label{parcorrelator}
{\cal D}(t^2)=
{b_0 \over t^4}\exp{(-t/\lambda_a)}+A_0\exp{(-t/\lambda_A)}
\qquad
{\cal D}_1(t^2)=
{b_1 \over t^4}\exp{(-t/\lambda_a)}+A_1\exp{(-t/\lambda_A)}
\,.
\ee
One would expect the $1/t^4$ terms to have something to do with the
perturbative terms. In any case, at short distances, the gluonic
correlator behavior should go closer to the one expected by
perturbation theory. Unfortunately, we see no indication of this but rather the
opposite. Whereas the perturbative result predicts a behavior $\sim -
1/t^4$ for the $\langle {\bf E}(t)\cdot {\bf E}(0) \rangle$ operator, the
lattice simulations get a {\it positive} slope at short distance (either for quenched or
unquenched simulations). Lacking of any 
explanation for this fact, we will refrain of using those results in
order to get an estimate of the non-perturbative effects (another point
of concern is that the gluon condensate prediction for quenched
simulations is one order of magnitude larger than the phenomenological
value). Therefore, we will not try to give any number
for the ultrasoft contribution in this paper and add them to the errors\footnote{In
principle, a similar problem may appear in sum rules calculation. See
the discussion at the end of this section.}. In order not to 
overestimate them, we will constraint the allowed range of values for
the ultrasoft contribution by some consistency arguments. 
If we rely on
power counting rules, since $\lQ \sim m\als^2$, the non-perturbative
effects would formally be of NNLO and from the discussion above about
the perturbative NNLO effects one would assume a value around $\sim 50$
MeV. On the other hand, for consistency of the theory ($\lQ^3/(m\als)^2
\ll m\als^2$), the ultrasoft corrections should be
smaller than the leading order solution (of order $\sim 160$ MeV). Note that if this 
consistency argument is not fulfilled, the very same assumption that
we can describe, in first approximation, the $\Upsilon(1S)$ by a
Coulomb-type bound state fails. 
In order to keep ourselves as
conservative as possible, we would only demand the ultrasoft corrections to be
smaller than the binding energy. Therefore, we will assign to our
evaluation of the RS mass a $\pm50$ MeV error. This is roughly
equivalent to assign a $\pm100$ MeV error to the evaluation of the
binding energy from the
ultrasoft contributions. This error may also be considered compatible with the
discussion of the situation where $m\als^2 \gg \lQ$. From the previous discussion about the magnitude of
the NNNLO contributions, we will add another $\pm25$ MeV error to our
evaluation of the RS mass (roughly equivalent to assign a $\pm50$ MeV error to the evaluation of the
binding energy) from any other (perturbative) source of higher order
effects. This may seem conservative in view of the estimates of the NNNLO
effects shown in Figs. \ref{figRSLL} and \ref{figRSprimeLL} and the
related discussion but we also include the
expected error from finite
charm mass effects (see \cite{Hoangcharm1}). Therefore, from the RS
scheme evaluation, our final estimate for the bottom mass
reads (in order to avoid double counting with the error from higher order effects, we do
not include now the scale dependence error)
\be
\label{MRSdet1}
m_{b,\RS}(2\;{\rm GeV})=4\,387^{+75}_{-75}({\rm US+higher\; orders})
^{-16}_{+16}(\als)^{-73}_{+75}(N_m+\nu_f)\;{\rm MeV};
\ee
\be
\label{MMSRSdet1}
m_{b,\MS}(m_{b,\MS})=4\,203^{+67}_{-67}({\rm US+higher\;
orders})^{-28}_{+27}(\als)^{-15}_{+16}(N_m+\nu_f)\;{\rm MeV}.
\ee
Whereas from the RS' scheme evaluation, we obtain the result (with the same variation of the
parameters)
 \be
\label{MRSprimedet1}
m_{b,\RS'}(2\;{\rm GeV})=4\,782^{+75}_{-75}({\rm US+higher\; orders})
^{+15}_{-12}(\als)^{-35}_{+31}(N_m+\nu_f)\;{\rm MeV};
\ee
\be
\label{MMSRSprimedet1}
m_{b,\MS}(m_{b,\MS})=4\,214^{+67}_{-67}({\rm US+higher\;
orders})^{-25}_{+25} (\als)^{-15}_{+12}(N_m+\nu_f)\;{\rm MeV}.
\ee

We average the two values obtained for the $\MS$ mass. We then obtain (rounding)
\be
\label{MMSRSprimedet2}
m_{b,\MS}(m_{b,\MS})=4\,210^{+90}_{-90}({\rm
theory})^{-25}_{+25}(\als)\;{\rm MeV}\,, 
\ee
where we have added (conservatively) the theoretical errors
linearly. These figures compare favorably (within errors) with other
determinations of the bottom mass. Either with other determinations
using the $\Upsilon(1S)$ mass (the second references in
\cite{sumrules2,EJBSV} and ref. \cite{Hoangcharm1}), sum rules \cite{sumrules2,Hoangcharm1},
lattice \cite{latticeb} or from measurements at LEP \cite{German}.

\medskip

In this paper, we have just taken into account the first IR renormalon
of the pole mass and the singlet static potential. Nevertheless, there are
subdominant renormalons that eventually could play a role. On the
singlet static
potential side, one expects the first problems to come from a
$O(\lQ^3r^2)$ IR renormalon. For the pole mass there are $O(\lQ^2/m)$
renormalons. A priori, it is not clear which one will play the dominant
role. In fact, it will depend on the relative size between
$\lQ$ and $m\als^2$. 

In the situation where a description in terms of local condensates is
appropriated ($m\als^2 \gg \lQ$), the leading {\it genuine}
non-perturbative corrections to the mass scale like $m(\lQ/m\als)^4$
but this quantity is (parametrically) much smaller than the
non-perturbative effects associated to the subleading renormalons, 
either from the pole mass or from the singlet static potential. Moreover,
in this case, parametrically, the leading ambiguity would come from
the subleading pole mass renormalon of $O(\lQ^2/m)$. Therefore, we
would be in a similar situation that when working with the pole mass,
where the actual accuracy of the result was set by the perturbative
calculation. Thus, one should first get rid of these subleading
renormalons in order to improve the accuracy of the calculation before
doing any reference about {\it genuine} non-perturbative effects.

In the situation where $m\als^2 \sim \lQ$, the subleading renormalon
ambiguities from the singlet static potential and the pole mass are
parametrically of the same order than the
{\it genuine} non-perturbative corrections. The $\Upsilon(1S)$ seems to
live closer to this situation.

\medskip

It is usually claimed that the non-perturbative effects in sum rules
are smaller than in the $\Upsilon(1S)$ mass. We would like to mention
that, at least parametrically, this is not the case under the standard
counting ${1 \over \sqrt{n}} \sim \als$ (where $n$ labels the moments
in sum rules, see \cite{sumrules1,sumrules2} for details). Nevertheless, it may
happen that they are numerically suppressed. This is indeed the case
if one considers that one can describe the non-perturbative effects by
local condensates \cite{Voloshin2}.  However, one can only use the
expression in terms of local condensates if one is in the situation
where ${m \over n} \gg \lQ$. This would be analogous to the assumption
$m\als^2 \gg \lQ$ and it may difficult to fulfill. Therefore, it is
more likely that the non-perturbative corrections will also depend on a
nonlocal condensate, in fact, arising from the same effective theory,
on the same chromoelectric correlator that the $\Upsilon(1S)$ mass
does. Thus, in order to estimate the non-perturbative errors in sum
rules evaluations, it would be most welcome to have, at least, the
explicit expression of the non-perturbative effects in the situation
${m \over n} \sim \lQ$, which, by now, is lacking. In that way, one
could relate the non-perturbative effects for different moments in the
sum rules or with the non-perturbative effects of the $\Upsilon(1S)$
mass and eventually search for less sensitive non-perturbative physics
observables.

\section{Charm $\MS$ quark mass determination}
\label{secdetmc}

In this section, we give a determination of the $\MS$ (and RS) charm mass. In
principle, one could think of using the same procedure than in the
previous section for the $J/\Psi$ (or $\eta_c$). This would imply to believe that the $J/\Psi$ is a mainly perturbative
system. We prefer to avoid this assumption in this work and to obtain the charm mass in a different
way. Our, maybe weaker, assumption will be to assume that HQET can be used
for charm physics. We will search for observables that are weakly
sensitive to non-perturbative physics. The observable that we will choose is the difference
between the B and D meson mass. Here, it follows an
analogous discussion than in the introduction. Whereas both the B
meson mass and the D meson mass are $O(\lQ)$ sensitive, the difference
is $O(\lQ^2/m)$ sensitive. Nevertheless, this improvement is not
exploited if one writes this observable in terms of the pole
masses. Therefore, here, we will use the RS masses and our previous determination of the bottom
mass in order to obtain the charm mass. 

The (spin-averaged) mass difference between the B and D meson mass
reads
\be
\label{mcequation}
\langle M_B \rangle-\langle M_D \rangle=m_{b,\RS}-m_{c,\RS}
+\lambda_1\left({1 \over 2m_{b,\RS}}-{1 \over 2m_{c,\RS}}\right)+O(1/m_{\RS}^2)
\,,
\ee
where $\lambda_1$ can be related with the expectation value of the
kinetic energy in HQET and  
\be
\langle M_{B(D)} \rangle={M_{B(D)}+3M_{B^*(D^*)} \over 4}
\,.
\ee
The value of $\lambda_1$ is poorly known. We take the value
$\lambda_1=0.3\pm0.2$ (see \cite{Neubert} and references therein). We
can now obtain the value of the charm mass. We perform the evaluation in
both schemes, the RS and RS'. In order to estimate the
errors, we fix $\nu_f=1$ GeV and allow for a variation of $\als$, $N_m$,
$\lambda_1$ and $m_b$ as follows: $\als(M_z)=0.118\pm 0.003$,
$N_m=0.552\pm 0.0552$,  $\lambda_1=0.3\pm0.2$ and
$m_{b,\MS}=4\,203^{+67}_{-67}$ MeV (in the RS scheme) and
$m_{b,\MS}=4\,214^{+67}_{-67}$ MeV (in the RS' scheme). For the variation of
$\als$ and $N_m$, we use the correlated values of the bottom mass (that is why
we do not include these errors in the variation of the bottom mass). For the error 
due to $\nu_f$ see the discussion below. We obtain
\be
\label{MRScdet1}
m_{c,\RS}(1\;{\rm GeV})=1\,181^{+82}_{-84}(m_{b,\MS})
^{+4}_{-1}(\als)^{-50}_{+50}(N_m)^{-78}_{+65}(\lambda_1)\;{\rm MeV};
\ee
\be
\label{MMSRScdet1}
m_{c,\MS}(m_{c,\MS})=1\,206^{+66}_{-67}(m_{b,\MS})
^{+1}_{-0}(\als)^{+11}_{-13}(N_m)^{-62}_{+52}(\lambda_1)\;{\rm MeV},
\ee
and for the RS' scheme
 \be
\label{MRScprimedet1}
m_{c,\RS'}(1\;{\rm GeV})=1\,477^{+79}_{-80}(m_{b,\MS})^{+34}_{-27}
(\als)^{-19}_{+19}(N_m)^{-54}_{+48}(\lambda_1)\;{\rm MeV};
\ee
\be
\label{MMSRScprimedet1}
m_{c,\MS}(m_{c,\MS})=1\,207^{+65}_{-64}(m_{b,\MS})^{-7}_{+5}(\als)^{+13}_{-14}(N_m)^{-43}_{+39}(\lambda_1)\;{\rm MeV}.
\ee
We can check that the perturbative relation between the RS and $\MS$
 charm mass is indeed convergent. We obtain
\bea
m_{c,\RS}(1\;{\rm GeV})&=&1\,206-53+20+6+3=1\,181\;
{\rm MeV},
\\
\nn
m_{c,\RS'}(1\;{\rm GeV})&=&1\,207+205+46+13 + 6=1\,477\;{\rm MeV}
\,.
\eea

\medskip

In order to test our evaluation, since, for the charm quark, $\nu_f$
 stays close to the charm mass value (maybe jeopardizing the real
structure of the perturbative expansion), we will also perform the calculation
 in the OS scheme (note that in that case large logs may 
 appear). We use Eq. (\ref{mcequation}) with the replacement RS
 $\rightarrow$ OS. In order to achieve the renormalon cancellation, we rewrite
 both $m_{b,\OS}$ and $m_{c,\OS}$ in terms of the respective $\MS$
 masses and expand them in $\als(m_{c,\MS})$. In this way, we obtain the renormalon cancellation for
 $m_{b,\OS} - m_{c,\OS}$ but not for the $1/m$ terms. Therefore, for the
 latter, we use the $\MS$ masses as the expansion parameters. This
 effectively increases the magnitude of the $1/m$ terms but the other
 consistent option, to use the OS masses, is heavily affected by the
 renormalon (for instance $m_{c,\OS}=2\,263$ MeV for $m_{c,\MS}=1\,210$ MeV) introducing even larger errors and not reflecting
 the real size of the $1/m$ corrections. We allow for the same
 variation of $\als$ and $\lambda_1$ than above, whereas for
 the bottom mass we use $m_{b,\MS}=4\,210\pm90$ MeV. We obtain
\be
\label{MMScdet2}
m_{c,\MS}(m_{c,\MS})=1\,254^{+85}_{-84}(m_{b,\MS})^{+17}_{-12}(\als)^{-49}_{+45}(\lambda_1)
\;{\rm MeV}
\,.
\ee
We can check
 the convergence of the perturbative expansion of $m_{b,\OS}-m_{c,\OS}$ in terms
 of the $\MS$ masses in order to test the validity of this
 evaluation. We obtain (order by order in $\als(m_{c,\MS})$)
\be
m_{b,\OS}-m_{c,\OS}=2\,956+490-14-32+22=3\,423\;{\rm MeV}
\,.
\ee
For the last terms in the series the situation is not conclusive. One
could think that the expansion is reliable up to, maybe, a $\sim \pm 30$
MeV uncertainty. One source of error comes from the expansion in
$\als$ since two kind of logs arise:
$\ln{(m_b/\nu)}$ and $\ln{(m_c/\nu)}$, which can not be minimized at
the same time. In fact, for the choice $\nu=2m_{c,\MS}$, we obtain
$m_{c,\MS}(m_{c,\MS})= 1\,239$ MeV and the expansion seems to improve: 
$$
m_{b,\OS}-m_{c,\OS}=2\,971+345 +79+19 +10=3\,424\;{\rm MeV}
.$$ 

Let us consider further sources of error in the RS scheme
evaluations. For these, we are using the RS bottom masses at a quite low
$\nu_f=1$ GeV. This produces that the convergence in the conversion
from the $\MS$ to the RS and RS' masses for the bottom quark is
slower, in particular for the RS' mass (see Table
\ref{tabmassesnuf2}). One can then believe that
higher orders in the relation between the $\MS$ and the RS' masses
will add further positive contribution that in turn will increase the value of the $\MS$
charm mass bringing it closer to the $\MS$ evaluation. In
any case, they are compatible within errors. Therefore, we add a $\pm40(20)$ MeV error to our RS'(RS) evaluation from
the conversion from the $\MS$ to the RS'(RS) bottom quark mass. 
Another source of
error comes from the $1/m$ terms. We have two effects compensating each
other. With the RS scheme, the magnitude of the corrections in the relation between the
 RS and the $\MS$ bottom quark mass is smaller but at the price of worsening the $1/m$
expansion (this observation could also apply to the OS calculation). With the RS' scheme, the magnitude of the corrections in the relation between the
RS and  $\MS$ bottom quark mass is larger but the $1/m$ expansion improves
(as well as being less sensitive to $\lambda_1$). We estimate the
$O(1/m^2)$ corrections to be  of order $\sim 15(30)$ MeV for the
RS'(RS) evaluation and also add them to the errors. Overall, the
theoretical errors in the RS and RS' evaluations are basically
equivalent. Therefore, we, somewhat, weigh more the RS' scheme
evaluation, being less sensitive to $\lambda_1$. Our final
result for the $\MS$ charm mass reads (rounding)
\be
\label{MMScdetfinal}
m_{c,\MS}(m_{c,\MS})=1\,210^{+70}_{-70}({\rm theory})^{+65}_{-65}(m_{b,\MS})^{-45}_{+45}(\lambda_1)\; {\rm MeV},
\ee
where we have not included the errors in $\als$ being negligible
compared with the other sources of error.
This result can be compared with two recent evaluations where Charmonium
data was used \cite{EJBSV}.

\medskip

At this stage, we can also give a prediction for
$\bar \Lambda$ by using 
\be
\label{lambdabarequation}
\bar \Lambda_{\RS} 
=
\langle M_B \rangle-m_{b,\RS}-{\lambda_1  \over 2m_{b,\RS}}+O(1/m_{b,\RS}^2)
\,.
\ee
We obtain (using $m_{b,\MS}=4\,210$ MeV)
\be
\bar \Lambda_{\RS}(1\;{\rm GeV}) = 659\;{\rm MeV}, 
\qquad 
\bar \Lambda_{\RS'}(1\;{\rm GeV})= 401\;{\rm MeV}. 
\ee
We can see that it is crucial to specify the scheme in order to give a
meaningful prediction for $\bar \Lambda$. 

\medskip

Finally, we would like to mention that $m_{b,\RS}-m_{c,\RS}$ (and
$m_{b,\OS}-m_{c,\OS}$) suffers from renormalon ambiguities of
$O(\lQ^2/m)$ that cancel with the renormalon of
$\lambda_1/m$. Therefore, one could argue whether it makes any sense
to give a value of $\lambda_1$ without specifying how to handle the
$O(\lQ^2/m)$ renormalon in $m_{b,\RS}-m_{c,\RS}$. This could only be
explained if the ambiguity due to the renormalon is much smaller than
the {\it genuine} non-perturbative effects. If we take as an indication
the perturbative expansions found above, we may believe that any
ambiguity in the perturbative expansion is smaller than the {\it genuine}
non-perturbative effects. On the other hand, this ambiguity may explain
the spread of values one can find in the literature for $\lambda_1$.

\section{Conclusions and outlook}
\label{conclusions}
We have approximately computed the normalization constant of the first
   infrared renormalon of the pole mass (and the singlet static
   potential). Estimates of the higher order coefficients of the
   perturbative series relating the pole mass with the $\MS$ mass (and
   the singlet static potential with $\als$) have been obtained
   without relying on the large $\beta_0$ approximation. New,
   renormalon free, definitions of the mass and potential have been
   given, within an effective field theory perspective, by
   subtracting their closest singularities in the Borel plane. We
   have obtained the bottom $\MS$ quark mass from the $\Upsilon(1S)$
   mass, 
   and an estimate of the errors, within this new scheme. We have also
   obtained the charm $\MS$ mass by using the mass difference between
   the $B$ and $D$ mesons. Our final figures read

\be 
m_{b,\MS}(m_{b,\MS})=4\,210^{+90}_{-90}({\rm
   theory})^{-25}_{+25}(\als)\; {\rm MeV} 
\ee 
and 
\be
   m_{c,\MS}(m_{c,\MS})=1\,210^{+70}_{-70}({\rm
   theory})^{+65}_{-65}(m_{b,\MS})^{-45}_{+45}(\lambda_1)\:{\rm MeV}.
\ee

\medskip

Several lines of research may follow from our results.

The use
of conformal mapping could eventually lead to an improved convergent series
in the evaluations of $N_m$ and $N_V$.

It would be interesting to apply the RS scheme to
bottomonium sum rules or to $t$-$\bar t$ production near threshold and
see how large the differences are with respect other determinations
avaliable in the literature. 

Our result for the bottom mass has been obtained in the zero mass
charm approximation. Finite charm mass effects \cite{Hoangcharm1,charm}
should be incorporated in future studies.

We have seen that our evaluation suffers from scale dependence for small
$\nu$. At NNLO, the main source of scale dependence comes from the
relativistic corrections. The incorporation of NNNLO effects does not
seem to correct this fact. It may happen that a renormalization-group
improved result could solve this problem. It is worth noting that
there already exists one result in the OS scheme in the situation where
$m\als^2 \gg \lQ$ \cite{HMS}. It would be desirable to have an
independent evaluation within pNRQCD and without relying on the
unequality $m\als^2 \gg \lQ$. After that, one should transform the
renormalization-group improved results from the OS to a renormalon free
scheme. This, a priori, may not turn out to be completely trivial.

One of the (potentially) major source of errors in our evaluation of the
bottom mass is the non-perturbative contribution. Any (reliable)
determination of this contribution will have an immediate impact on our
understanding of the theoretical errors. On the one hand, it would put
on more solid basis our implicit assumption that the leading order
solution corresponds to a Coulomb-type bound state and, once this is
achieved, it would move the error estimates of the non-perturbative
effects from a qualitative level to a quantitative one, 
(hopefully) bringing them down significantly.  On the other hand, one may
think of cross-checking our result with other determinations. The
fact that the difference happens to be relatively tiny supports our
believe that (perturbative and non-perturbative) higher order effects are
indeed not very large. Alternatively, one may search for combinations of
observables less sensitive to long distance physics effects in order to
get a more accurate result for the masses.

Leaving aside theoretical errors, one may expect to 
bring down the errors associated to other parameters of the theory
significantly (if
one has a large enough set of observables) by using global fits.

Another issue that deserves further consideration is whether it is
possible to develop a renormalon subtraction scheme completely within
dimensional regularization (see the discussion at the beginning of
sec. \ref{secdefRS}). This would provide a better understanding of the
physical system and therefore the errors could be estimated in a more
reliable way.

We expect to come back to these issues in the near future.

\medskip

{\bf Acknowledgments} 

I thank K.G. Chetyrkin and A. Hoang for useful discussions and
N. Brambilla and A. Vairo for showing me some of their pictures prior to
publication.

\end{document}